\documentclass[final,5p,times,twocolumn,authoryear]{elsarticle}

\usepackage{graphicx}
\usepackage{amssymb}
\usepackage{amsmath}
\usepackage{amsfonts}
\usepackage{natbib}
\usepackage{rotating}
\usepackage[dvipsnames]{xcolor}

\journal{Icarus}

\begin{document}

%\linenumbers

\begin{frontmatter}

\title{SPH simulations of high-speed collisions between asteroids and comets}

\author{J.~Rozehnal}
\author{M.~Bro\v z}

\address{Charles University, Faculty of Mathematics and Physics, Institute of Astronomy, V Hole\v sovi\v ck\'ach 2, 18000 Prague 8, Czech Republic}

\author{D.~Nesvorn\'y}
\author{K.~J.~Walsh}
\author{D.~D.~Durda}

\address{Southwest Research Institute, 1050 Walnut Street, Suite 400, Boulder, CO 80302, USA}

\author{D.~C.~Richardson}

\address{Department of Astronomy, University of Maryland, College Park, MD 20742, USA}

\author{E.~Asphaug}

\address{Lunar and Planetary Laboratory, University of Arizona, Tucson, Arizona, USA}

%%%%%%%%%%%%%%%%%%%%%%%%%%%%%%%%%%%%%%%%%%%%%%%%%%%%%%%%%%%%%%%%%%%%%%%%

\begin{abstract}
We studied impact processes by means of smoothed-particle hydrodynamics (SPH)
simulations. The method was applied to modeling formation of main-belt families
during the cometary bombardment (either early or late, ${\sim}\,3.85\,{\rm Gy}$
ago). If asteroids were bombarded by comets, as predicted by the Nice model,
hundreds of asteroid families (catastrophic disruptions of diameter $D \ge 100\,{\rm
km}$ bodies) should have been created, but the observed number is only 20.
Therefore we computed a~standard set of 125 simulations of collisions between
representative $D = 100\,{\rm km}$ asteroids and high-speed icy projectiles (comets),
in the range 8~to 15\,km/s.
According to our results, the largest remnant mass~$M_{\rm lr}$ is similar as in low-speed collisions,
due to appropriate scaling with the effective strength~$Q_{\rm eff}$,
but the largest fragment mass~$M_{\rm lf}$ exhibits systematic differences
--- it is typically smaller for craterings and bigger for super-catastrophic events.
This trend does not, however, explain the non-existence of old families.
The respective parametric relations can be used in other statistical (Monte-Carlo) models
to better understand collisions between asteroidal and cometary populations.
\end{abstract}


\end{frontmatter}

% \linenumbers

%%%%%%%%%%%%%%%%%%%%%%%%%%%%%%%%%%%%%%%%%%%%%%%%%%%%%%%%%%%%%%%%%%%%%%%%

\section{Introduction}\label{sec:introduction}

In our previous work \cite{Broz_2013A&A...551A.117B}, we asked a question whether some of main-belt asteroid families had been formed by collisions with 
(originally trans-Neptunian) comets during the period of the late heavy bombardment (LHB), as predicted by the Nice model \citep{Gomes_2005Natur.435..466G,Morbidelli_2005Natur.435..462M},
or its newer variants \citep{Morbidelli_2010AJ....140.1391M,Nesvorny_2012AJ....144..117N}.

We concluded that if asteroid families were created during the LHB,
the final number of catastrophic disruptions with parent bodies larger than
$D_{\rm{pb}} \simeq 100\,\,\rm{km}$ in diameter should be ${\sim}\,100$,
i.e., almost one order of magnitude larger than the observed number ($20$).
Also, the synthetic production function, i.e., the cumulative number of families
with the parent body size larger than~$D$,
is significantly steeper in the collisional model than observed.

There are three possible explanations for this discrepancy (apart from secondary disruptions of family members, which certainly contributes to the decrease of kilometre-sized bodies):
\begin{itemize}
\item[(i)] cometary flux could have been reduced 
by 80\,\% due to intrinsic activity and breakups of the cometary nuclei during their close approaches to the Sun;
\item[(ii)] physical lifetime of the comets may be size-dependent, so that 
small comets are disintegrated substantially more than the large ones
(see the discussion of their size-frequency distribution in \citealt{Broz_2013A&A...551A.117B});
\item[(iii)] collisions between solid monolithic targets (asteroids) and less cohesive projectiles 
(comets) occurring at high speeds ($v_{\rm imp} > 10\,{\rm km}\,{\rm s}^{-1}$) may be generally different from collisions at low speeds, which are the only cases that have been studied so far \citep{Benz_1999Icar..142....5B,Durda_2007Icar..186..498D,Benavidez_2012Icar..219...57B,Benavidez_2018Icar..304..143B,Jutzi_2015P&SS..107....3J,Sevecek_2017Icar..296..239S,Sevecek_2019A&A...629A.122S}.
\end{itemize}
In this work, we focus on the latter possibility.

We model mutual collisions of asteroids with cometary nuclei occurring at high relative velocities \citep{Vokrouhlicky_2008AJ....136.1463V,Broz_2013A&A...551A.117B}.
We simulate impacts of icy, low-density ($\rho_{\rm bulk} = 1.14\,{\rm g}\,{\rm cm}^{-3}$) projectiles with basaltic monolithic targets at velocities of $8$~to $15\,{\rm km}\,{\rm s}^{-1}$.
We focus on possible differences in the propagation of the shock wave, ejection of fragments and resulting differences in the size-frequency distribution (SFD)
of synthetic asteroid families. We compare our results with simulations of mutual collision of basalt bodies, occurring at lower speeds (3~to $7\,{\rm km}\,{\rm s}^{-1}$),
typical for the main asteroid belt.
We also discuss a scaling of SFDs with respect to the `nominal' target diameter $D = 100\,\,{\rm km}$, for which a number of simulations have been done so far 
(e.g., \citealt{Durda_2007Icar..186..498D,Benavidez_2012Icar..219...57B,Benavidez_2018Icar..304..143B,Jutzi_2015P&SS..107....3J,Sevecek_2017Icar..296..239S,Sevecek_2019A&A...629A.122S}).

%%%%%%%%%%%%%%%%%%%%%%%%%%%%%%%%%%%%%%%%%%%%%%%%%%%%%%%%%%%%%%%%%%%

\section{Methods}\label{sec:methodsch4}

\def\d{{\rm d}}
\def\vec#1{{\bf #1}}

The method we chose for the simulations of collisions between a solid bodies is a meshless Lagrangian particle method -- smoothed--particle hydrodynamics (SPH) (e.g., \cite{Benz_1990nmns.work..269B,Monaghan_1992ARA&A..30..543M,Benz_1994Icar..107...98B}). 
Gas and solid bodies are modelled by the respective set of partial differential equations, which are summarized as follows:

\begin{equation}
{\d\rho\over\d t} = -\rho\nabla\cdot\vec v\,,\label{drho_dt}
\end{equation}
\begin{equation}
{\d\vec v\over\d t} = -{1\over\rho}\nabla P - \nabla\Phi + {1\over\rho}\nabla\cdot\vec S\,,
\end{equation}
\begin{equation}
{\d U\over\d t} = -{1\over\rho}P\nabla\cdot{\vec v} + {1\over\rho}{1\over 2}\left[\nabla\vec v + (\nabla\vec v)^{\rm T}\right]:\vec S\,,
\end{equation}
\begin{equation}
\nabla\cdot\nabla\Phi = 4\pi G\rho\,,
\end{equation}
\begin{equation}
{\d\vec S\over\d t} = \nabla\cdot\mu_1\left[\nabla\vec v + (\nabla\vec v)^{\rm T}\right] + \left(\mu_2-{2\over 3}\mu_1\right)\nabla\cdot \vec v\vec I\,,
\end{equation}
\begin{equation}
{\d D\over\d t} = \left({c_{\rm g}\over R_{\rm s}}\right)^3 + \left({m+3\over 3}\alpha^{1\over 3}\epsilon^{m\over 3}\right)^3\,,
\end{equation}
\begin{equation}
P \simeq A\left({\rho\over\rho_0}-1\right) + B\left({\rho\over\rho_0}-1\right)^2 + b\rho U {1\over {U\over U_0}{\rho_0^2\over\rho^2}+1} + a\rho U\,,\label{tillotson}
\end{equation}
\begin{equation}
\vec S := (1-D)\, {\rm min}\left[{Y^2\over {3\over 2}\vec S:\vec S}; 1\right] \vec S\,,
\end{equation}
\begin{equation}
P := {\cal H}(P) + {\cal H}(-P)(1-D)P\,,\label{P}
\end{equation}
where
$\rho$ denotes the bulk density,
$\vec v$ velocity,
$P$ pressure,
$\Phi$ gravitational potential,
$\vec S$ stress tensor,
$U$ specific internal energy (per unit mass),
$D$ scalar damage,
$\mu_1$ shear modulus,
$\mu_2$ bulk modulus,
${\bf I}$ unit matrix,
$Y$ von Mises limit,
$c_{\rm g}$ crack propagation speed,
$R_{\rm s}$ equivalent particle size,
$\alpha \equiv 8\pi c_{\rm g}^3 k/[(m+1)(m+2)(m+3)] $,
$k$, $m$ parameters of the Weibull distribution,
$n(\epsilon) = k\epsilon^m$,
$\epsilon$ flaw activation limit,
${\cal H}$ the Heaviside step function.
The individual terms (right-hand sides) can be briefly
described as follows:
expansion,
pressure gradient,
gravity,
stress,
work,
viscous heating,
Poisson,
shear stress,
bulk stress,
crack growth,
activation,
solid-state pressure,
quadratic term,
corrective term,
ideal-gas term,
fracture,
plasticity
for the stress and for the pressure
($:$~means double-dot product, $:=$~assignment).
In the SPH5 \citep{Benz_1994Icar..107...98B} code we used, the spatial discretisation of Eqs.~(\ref{drho_dt})--(\ref{P})
is performed in the standard SPH way
\citep{Monaghan_1992ARA&A..30..543M,Benz_1994Icar..107...98B},
including the artificial viscosity to handle shocks.
For the temporal discretisation, the predictor-corrector method
is used (or alternatively, we implemented the Bulirsch-Stoer).

We assumed the Tillotson equation of state (Eq.~(\ref{tillotson});
\citealt{Tillotson_1962geat.rept.3216T}) and material properties,
which are listed in Table~\ref{tab:materials}. Let us note that
the Tillotson EOS lacks a melt phase, and indeed is not thermodynamically
consistent as it lacks latent heats: however, it is well suited
to simulations where material is mostly either in solid state
or the expanded state, which is our case.

\begin{table}\vspace{-5mm}\centering
\caption{Material constants used in our SPH simulations for basalt and
silicated ice (\hbox{30\,\%} of silicates).
Listed here are:
the zero-pressure density $\rho_0$,
bulk modulus $A$,
non-linear compressive term $B$,
sublimation energy $E_0$,
Tillotson parameters $a$, $b$, $\alpha$ and $\beta$,
specific energy of incipient vaporization $E_{\rm iv}$,
complete vaporization $E_{\rm cv}$,
shear modulus $\mu$,
plastic yielding $Y$,
melt energy $E_{\rm melt}$
and Weibull fracture parameters $k$ and $m$.
Values we used for silicated ice are identical to those of pure ice, except
density $\rho_0$, bulk modulus $A$ and Weibull parameters $k$ and $m$.
All values were adopted from \cite{Benz_1999Icar..142....5B}.
}
\vspace{2mm}
\small
\renewcommand{\tabcolsep}{6pt}
\begin{tabular}{|c|c|c|c|}
\hline\
quantity & basalt & silicated ice & unit\\
\hline\hline
$\rho_0$ & 2.7 & 1.1 & ${\rm g}\,{\rm cm}^{-3}$ \\
$A$ & $2.67\cdot10^{11}$ & $8.44\cdot10^{10}$ & ${\rm erg}\,{\rm cm}^{-3}$ \\
$B$ & $2.67\cdot10^{11}$ & $1.33\cdot10^{11}$ & ${\rm erg}\,{\rm cm}^{-3}$ \\
$E_0$ & $4.87\cdot10^{12}$ & $1.00\cdot10^{11}$ & ${\rm erg}\,{\rm g}^{-1}$ \\
$a$ & 0.5 & 0.3 & -- \\
$b$ & 1.5 & 0.1 & -- \\
$\alpha$ & 5.0 & 10.0 & -- \\
$\beta$ & 5.0 & 5.0 & -- \\
$E_{\rm iv}$ & $4.72\cdot10^{10}$ & $7.73\cdot10^{9}$ & ${\rm erg}\,{\rm 
g}^{-1}$ \\
$E_{\rm cv}$ & $1.82\cdot10^{11}$ & $3.04\cdot10^{10}$ & ${\rm erg}\,{\rm 
g}^{-1}$ \\
$\mu$ & $2.27\cdot10^{11}$ & $2.80\cdot10^{10}$ & ${\rm erg}\,{\rm cm}^{-3}$ \\
$Y$ & $3.5\cdot10^{10}$ & $1.0\cdot10^{10}$ & ${\rm erg}\,{\rm g}^{-1}$ \\
$E_{\rm melt}$ & $3.4\cdot10^{10}$ & $7.0\cdot10^{9}$ & ${\rm erg}\,{\rm 
g}^{-1}$ \\
$k$ & $4.0\cdot10^{29}$ & $5.6\cdot10^{38}$ & ${\rm cm}^{-3}$ \\
$m$ & 9.0 & 9.4 & -- \\
\hline
%\vspace{2mm} 
\end{tabular}
\label{tab:materials}
\end{table}

We performed 125 simulations of impacts of various projectiles
on targets with diameters $D_{\rm{pb}} = 100\,\rm{km}$.
The projectile velocity $v_{\rm{imp}}$ was 8, 10, 12, 14 and $15\,{\rm km}\,{\rm s}^{-1}$ and the impact angle $\varphi_{\rm{imp}}$ was $15^\circ, 30^\circ, 45^\circ, 60^\circ\, \rm{and}\, 75^\circ $.

The target was always basalt with 
bulk density $\rho_{\rm t} = 2.7\,{\rm g}\,{\rm cm}^{-3}$, while the projectiles consisted of silicated ice (\hbox{30\,\%} silicates) with bulk density ($\rho_{\rm imp} = 1.14\,{\rm g}\,{\rm cm}^{-3}$). 

The integration was controlled by the Courant number $C = 1.0$, a typical 
time step thus was $\Delta t \simeq 10^{-5}\,{\rm s}$, and the time span was 
$t_{\rm stop} = 100\,{\rm s}$. The Courant condition was the same in different 
materials, using always the maximum sound speed $c_{\rm s}$ among
all SPH particles, as usually. 

We used $N_{\rm SPH, t}=10^5$~SPH particles for 
the single spherical target. The impactor was modelled by $N_{\rm SPH, i}=10^3$ SPH particles.

We terminated SPH simulations after 50\,s from the impact. This time 
interval is sufficient to complete the fragmentation and establish a velocity field of fragments in our set of high-speed simulations.
Then we handed the output of the SPH simulation as initial 
conditions to the N--body gravitational code PKDGRAV \citep{Richardson_2000Icar..143...45R}, 
which is a parallel tree code used to simulate a gravitational reaccumulation
of fragments. We calculated fragments radii from their masses~$m$
and densities~$\rho$ as:
\begin{equation}
R = \left({3m\over 4\pi\rho}\right)^{1/3}\,.\label{R_m}
\end{equation}

We ran the N-body simulation with a time step $\Delta t = 5.0\,{\rm s}$ and we terminated it
after $t_{\rm evol} = 3$~days of evolution. To ensure 
this is sufficiently long, we also ran several simulations with $t_{\rm evol} = 
5$~days, but we saw no significant differences between final results.
For simplicity, we assumed perfect merging.
Particles are merged whenever they collide (or overlap),
regardless of their kinetic or rotational energy.

We used the nominal value for the tree opening angle,
${\rm d}\theta = 0.5\,{\rm rad}$, even though for the
evolution of eventual moons it would be worth using an even
smaller value, e.g., ${\rm d}\theta = 0.2\,{\rm rad}$. 

To compare the resulting SFDs properly, we varied the mass (and thus the size) of the projectile to obtain the same ratio of impactor specific energy $Q$ to the target strength $Q^*_{\rm{D}}$
within each simulation set.

%%%%%%%%%%%%%%%%%%%%%%%%%%%%%%%%%%%%%%%%%%%%%%%%%%%%%%%%%%%%%%%%%%%

\section{Results}\label{sec:results}

\begin{figure*}
\centering
\includegraphics[width=17.5cm]{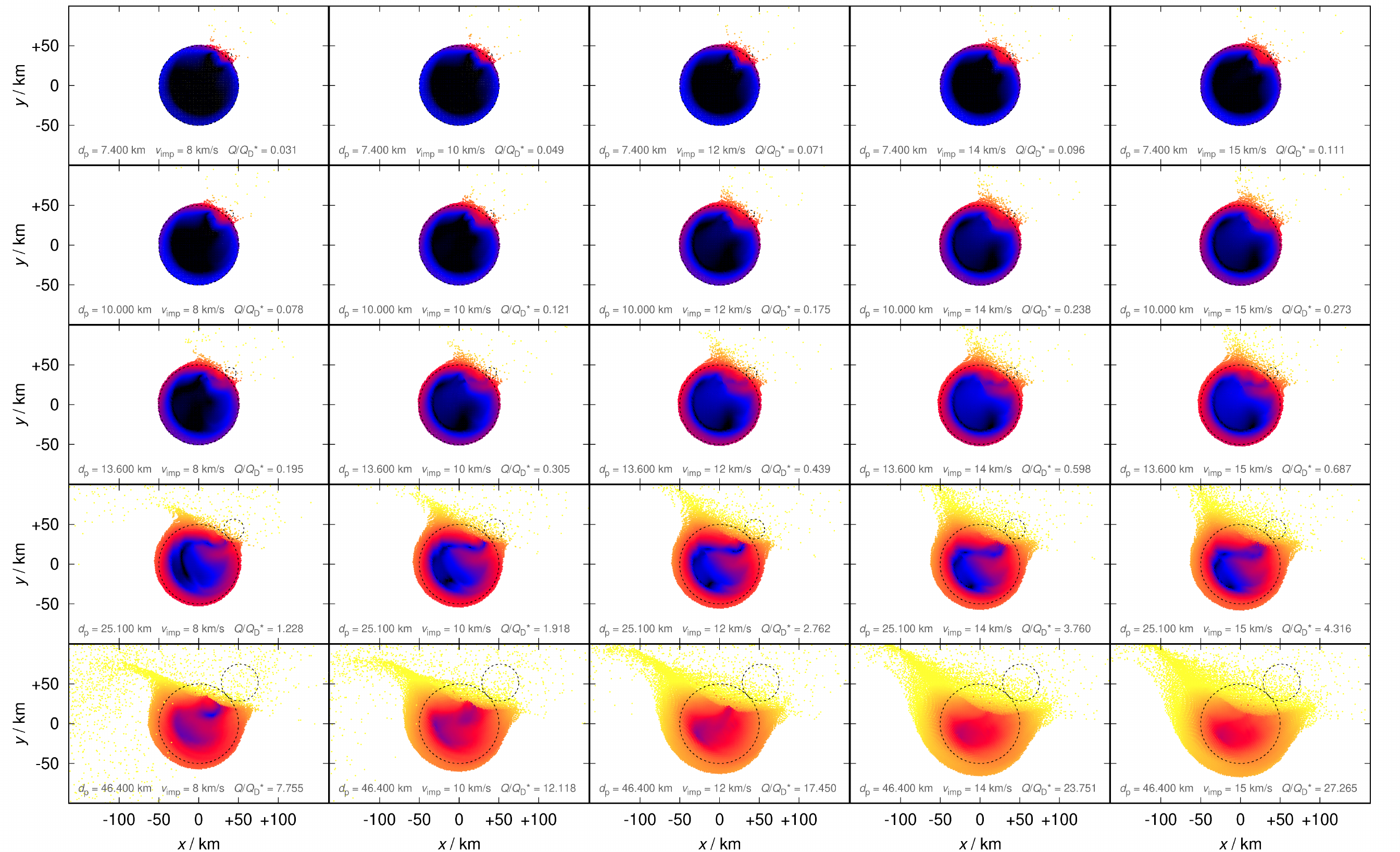}
\caption{Spatial distribution of SPH particles
at the end of the fragmentation phase (i.e., time 50\,s).
The interior of the target is shown,
with the $(x, y)$ cross-section and $z < 0$.
Dashed circles correspond to the initial positions
of the target and projectile.
$d_{\rm p}$ denotes the projectile size,
$v_{\rm imp}$ the impact speed,
$Q$~the strength,
$Q^\star_{\rm D}$ the scaling law.
Colours correspond to the logarithm of velocity (in m/s):
0~(black),
1~\color{blue}blue\color{black},
2~\color{red}red\color{black},
3~\color{Goldenrod}yellow\color{black}.
}
\label{xy_45}
\end{figure*}

\begin{figure*}
\centering
\includegraphics[width=17.5cm]{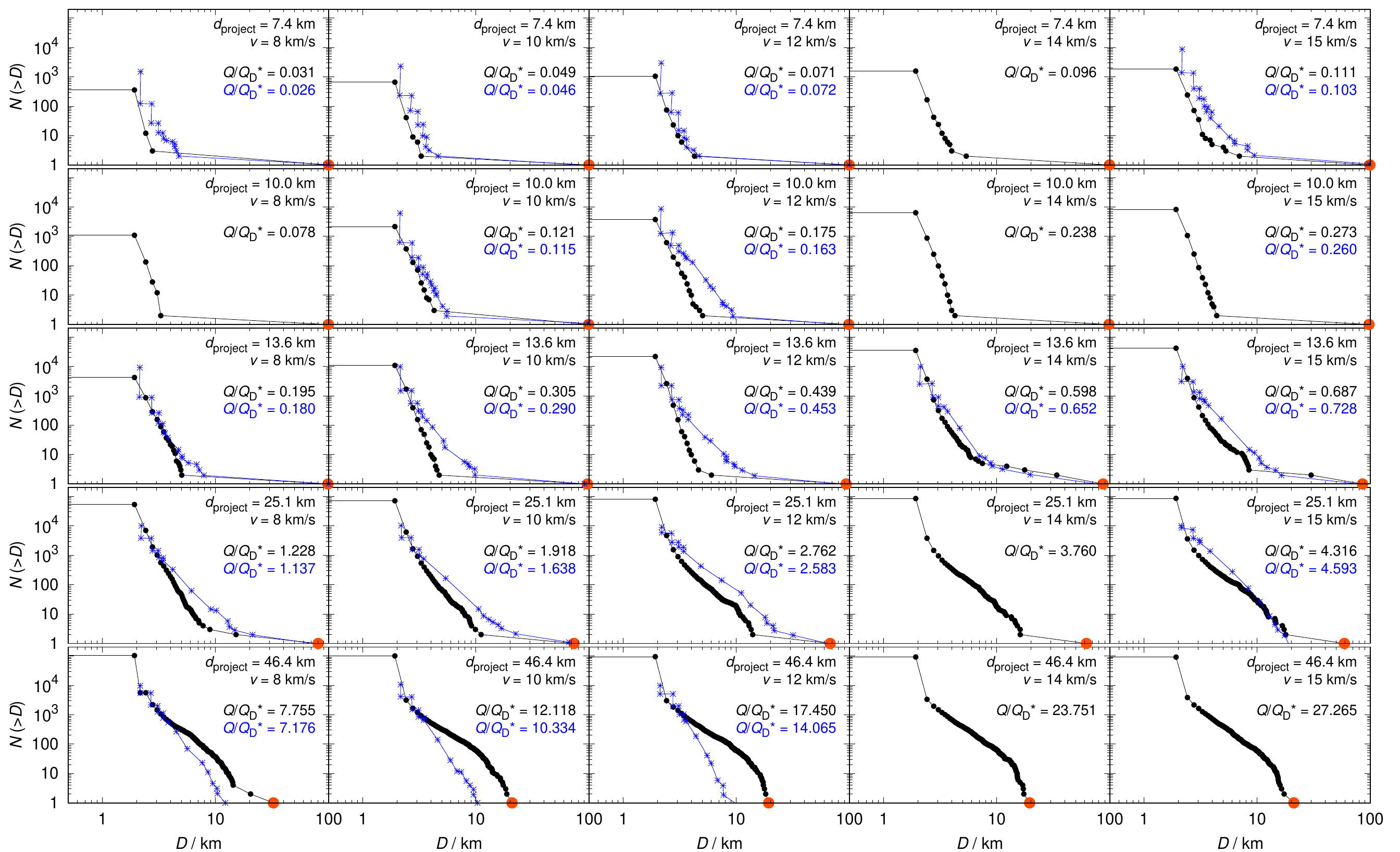}
\caption{Size-frequency distributions $N({>}D)$
for the impact angle $45^\circ$
and various projectile sizes $d_{\rm p}$ and impact velocities~$v_{\rm imp}$
(black).
The largest remnant (or the largest fragment) is indicated (\color{OrangeRed}red\color{black})
A comparison with \cite{Durda_2007Icar..186..498D}
at a similar specific energy $Q/Q^\star_{\rm D}$
is plotted (\color{blue}blue\color{black}), when available.}
\label{size_distribution_45_DURDA}
\end{figure*}

\begin{figure*}
\centering
\includegraphics[width=17.5cm]{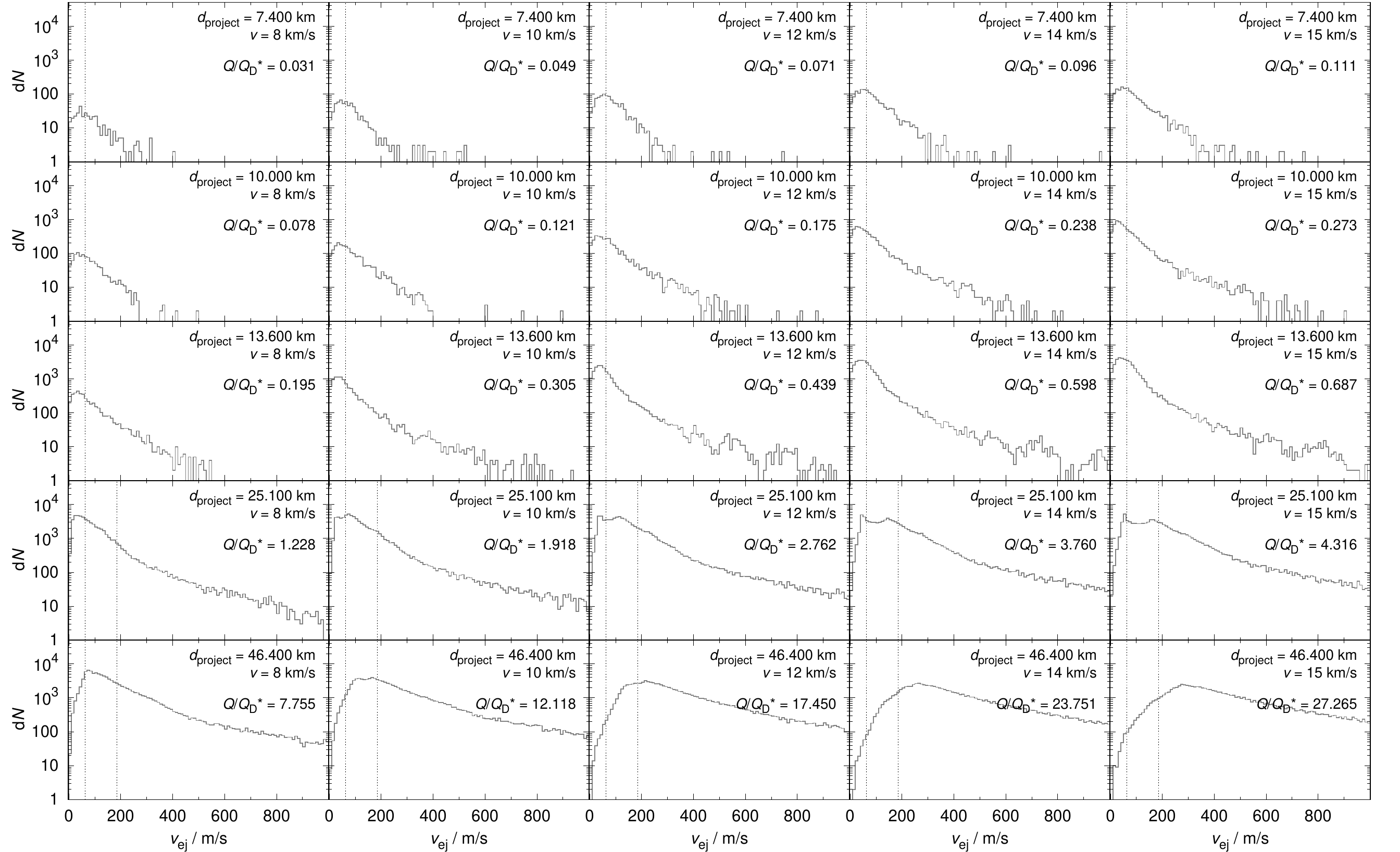}
\caption{Histograms of ejection velocities $\d N(v, v+\d v)$
for the impact angle $45^\circ$
and various projectile sizes and impact velocities.
The velocities were transformed to the barycentric frame,
with outliers ($v_{\rm ej} > 1000\,{\rm m}\,{\rm s}^{-1}$) removed.
The escape velocity ($v_{\rm esc} = 61\,{\rm m}\,{\rm s}^{-1}$)
is denoted by the vertical dotted line.
For high-energy impacts, we indicate a 2nd peak at approximately $3v_{\rm esc}$.
}
\label{hist_velocity_45}
\end{figure*}

\begin{figure*}
\centering
\includegraphics[width=17.5cm]{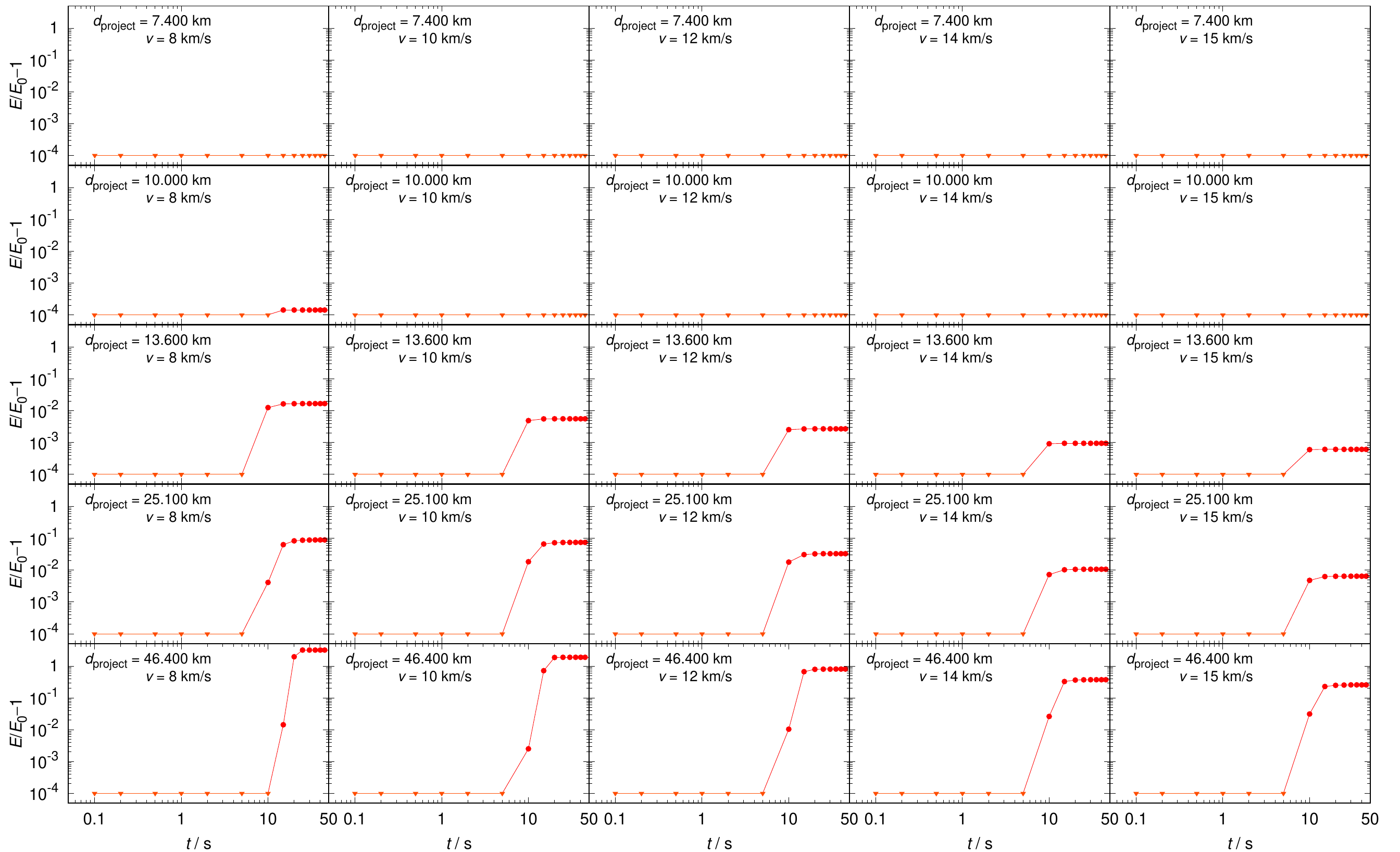}
\caption{Relative total energy $|E/E(t=0)-1|$ vs. time~$t$
for the impact angle $45^\circ$
and various projectile sizes and impact velocities.
If the relative energy difference is less than $10^{-4}$,
it is plotted at the ordinate (as a~triangle).}
\label{energy_45}
\end{figure*}

We processed the results of simulations and plotted
the spatial distribution at the end of the fragmentation phase (e.g.,~Fig.~\ref{xy_45} for impact angle $45^\circ$),
size-frequency distributions (Fig.~\ref{size_distribution_45_DURDA}),
velocity histograms (Fig.~\ref{hist_velocity_45}), and
energy vs.\ time (Fig.~\ref{energy_45}).

Impacts follow a regular pattern with increasing impact energy:
cratering $\rightarrow$
reaccumulative collision $\rightarrow$
catastrophic disruption $\rightarrow$
super-catastrophic disruption,
as one can see by comparing Figures \ref{xy_45}, \ref{xy_15} -- \ref{xy_75}.
For low-energy impacts, there is cratering only,
the bulk of the target remains with low ejection velocities (see Fig.~\ref{xy_45})
and its fragments are almost immediately reaccumulated;
the target is fully damaged though.
The impactor is vaporized, which is typical for low impact angles,
or dispersed by the reverse shock, for high impact angles
if its `upper' part misses the target.

The low-energy and high-impact-angle impacts are poorly resolved
(Fig.~\ref{xy_75}, top left),
and we should keep this in mind when interpreting the results.

\paragraph{Size-frequency distributions}
Constructing the SFDs might depend on the $R(m)$ relation
used to convert masses to radii.
At the end of the fragmentation phase, we used Eq.~(\ref{R_m}),
unlike \cite{Durda_2007Icar..186..498D}, who calculated radii
from the smoothing length $h$ as $R=h/3$.
This is only a minor difference though,
because both $\rho$ and $h$ are evolved in the course of time,
in accord with Eq.~(\ref{drho_dt}).
At the end of the reaccumulation phase, we used the original target density~$\rho_0$
to compute final diameters of fragments.
Regarding the largest remnant (LR), the largest fragment (LF) as well as other fragments,
they may be `puffed-up' in our simulations, if they have densities $\rho \ll \rho_0$.
This procedure is equivalent to assuming subsequent compaction of bodies (on long time scales).

Similarly as above, the SFDs follow a regular pattern with increasing impact energy.
The LR becomes smallish for high-energy impacts until it disappears;
these are super-catastrophic disruptions.
At the same time, the LF becomes bigger and bigger,
until the trend reverses and it becomes smaller for the highest-energy impacts.
The largest largest fragment (`LLF') reaches $D \simeq 20\,{\rm km}$ in our simulations
(see, e.g., Fig.~\ref{size_distribution_45_DURDA}, bottom right). 
A certain part of fragment size distribution,
with sizes ranging somewhere from $D \simeq 5\,{\rm km}$ up to $D \simeq 20\,{\rm km}$,
can be described by a power law $\log N({>}D) = c\log D + C$.
Typically it is steep, with slope $c \ll -2.5$,  
i.e., steeper than the collisional equilibrium of \cite{Dohnanyi_1969JGR....74.2531D},
but for super-catastrophic impacts this steep slope becomes shallow below $D \simeq 10\,{\rm km}$.
This is because these disruptions produce more fragments with diameters $D > 10\,{\rm km}$
and the total mass must be conserved ($M_{\rm tot} = M_{\rm pb} + m_{\rm imp}$).

\paragraph{Velocity distributions}
Regarding velocity distributions, we performed a transformation to the barycentre frame
after the reaccumulation phase.
To avoid a `contamination' of our fragment sample by the fragments of the impactor,
we removed outliers with ejection speed $v_{\rm ej} > 1000\,{\rm m}\,{\rm s}^{-1}$.

For low-energy impacts, there is a peak at about the escape velocity,
which is $v_{\rm esc} = 61\,{\rm m}\,{\rm s}^{-1}$ for our targets.
Practically all fragments are ejected within $2v_{\rm esc}$
(see Fig.~\ref{hist_velocity_45}; note the ordinate is logarithmic).

For high-energy impacts with a head-on geometry
(impact angle $\varphi_{\rm{imp}} \leq 45^{\circ}$),
a second peak appears at about $3v_{\rm esc}$,
probably due to a direct momentum transfer from the projectile to the target.
Eventually, the whole distribution shifts towards higher velocities (and the peaks merge).
These observations are very similar to those in \cite{Sevecek_2017Icar..296..239S}.

\paragraph{Energy conservation}
Let us note we experienced some problems with energy conservation.
In a majority of simulations the total energy (kinetic plus internal, target plus impactor)
is conserved to within $1\,\%$ (or better).
However, in a minority of simulations,
in particular the highest-energy and highest-impact-angle simulations,
the energy during the fragmentation phase is conserved to within $10\,\%$ (or even worse). 

Regarding high impact angles, we tracked-down this issue to projectile particles
exhibiting strong shearing very late in the fragmentation phase,
when the projectile is essentially `cut' by the target.
Practically, it does not affect the target at all,
because this `shearing instability' occurs elsewhere
(approximately ${\sim}\,300\,{\rm km}$ away from the impact site).

%%%%%%%%%%%%%%%%%%%%%%%%%%%%%%%%%%%%%%%%%%%%%%%%%%%%%%%%%%%%%%%%%%%

\paragraph{Parametric relations}
In order to describe the overall statistical properties
of collisions, it is useful to derive parametric relations,
which describe the dependence of
the largest remnant mass $M_{\rm lr}$,
the largest fragment mass $M_{\rm lf}$, or
the fragment size-distribution slope $q$
on a suitable measure of energy.

\begin{figure}
\centering
\includegraphics[width=8.5cm]{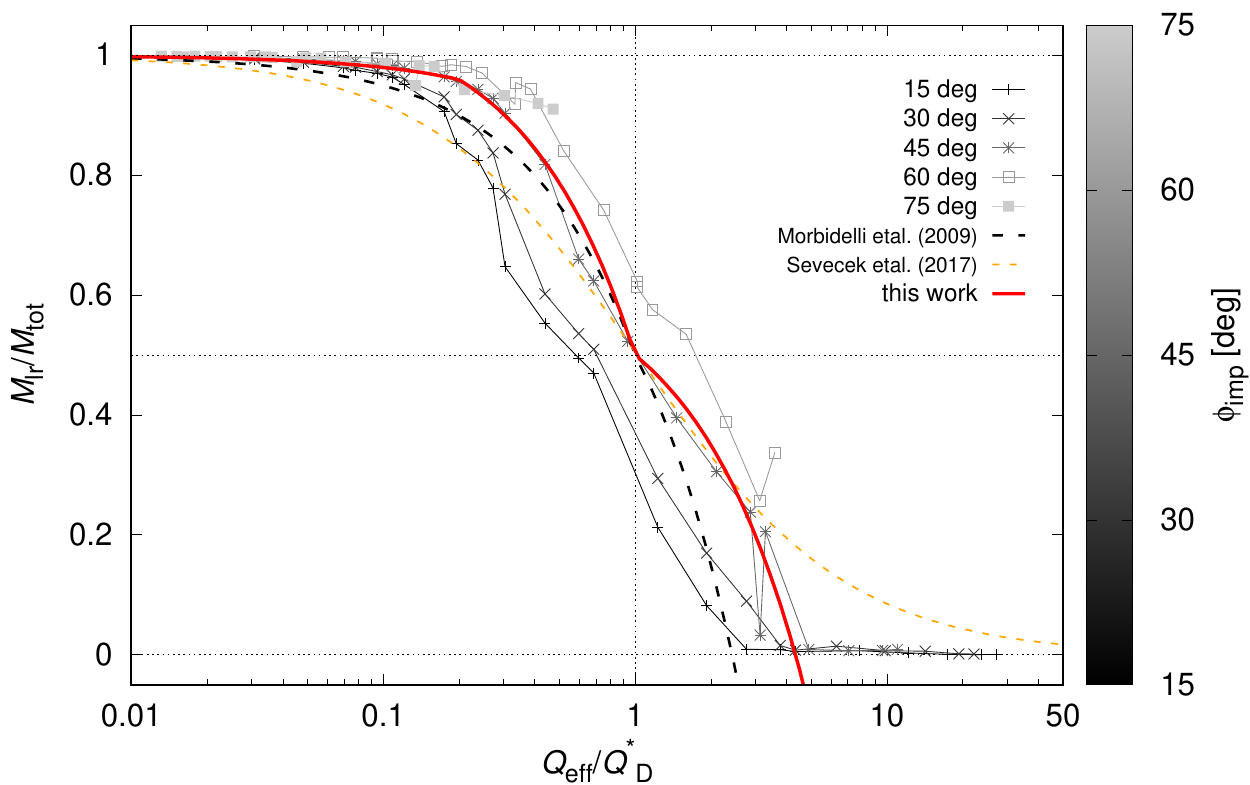}
\caption{Mass of the largest remnant~$M_{\rm lr}$ vs.\
the effective strength~$Q_{\rm eff}$.
Individual outcomes of our high-speed SPH simulations
are plotted as points.
The parametric relation Eq.~(\ref{Mlr}) is also plotted (solid red line).
In accord with the scaling law,
$M_{\rm lr}/M_{\rm tot} \simeq 0.5$ for $Q = 1$.
For comparison, we show the parametric relations
for low-speed collisions \citep{Morbidelli_2009Icar..204..558M}
and $D = 10\,{\rm km}$ bodies \citep{Sevecek_2017Icar..296..239S}.
}
\label{Mlr_Qeff}
\end{figure}

To this point, we use the standard scaling law
(as \citealt{Benz_1994Icar..107...98B}):
\setbox1=\hbox{\includegraphics[height=0.7em]{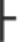}}
\setbox2=\hbox{\includegraphics[height=0.5em]{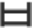}}
\begin{equation}
Q^\star_{\rm D}(R) = Q_0 R^{\kern1pt\copy1} + B\rho R^{\kern.7pt\copy2}\,,
\end{equation}
where $Q^\star_{\rm D}$ denotes the strength (in ${\rm erg}\,{\rm g}^{-1}$)
needed to disperse half of target,
$R$ its radius,
and $Q_0$, {\box1}, $B$, {\box2} parameters.
We used the same numerical values as \cite{Benz_1994Icar..107...98B},
because we used the same monolithic targets.
Moreover, we define the effective strength
(as \citealt{Sevecek_2017Icar..296..239S}):
\begin{equation}
Q_{\rm eff} = {{1\over 2}m v^2\over M_{\rm pb}} {A\over\pi r^2}\,,\label{Qeff}
\end{equation}
where the interacting cross section~$A$,
at the closest distance $d = (r+R)\sin(\phi)$, is:
\begin{equation}
A = \begin{cases}
\pi r^2 \quad\hbox{for } d \le R-r\,, \\
r^2\arccos{\frac{d^2+r^2-R^2}{2dr}} + R^2\arccos\frac{d^2+R^2-r^2}{2dR}\,- \\
\quad-\,{1\over 2}\sqrt{\textstyle(R+r-d)(d+r-R)(d-r+R)(d+r+R)} & \\
\quad\hbox{for } d\in(R-r; R+r) \,, \\
0 \quad\hbox{for } d \ge R+r\,. & \\
\end{cases}
\end{equation}
We verified that the high impact speeds do not substantially
alter the scaling law (see the point
$Q_{\rm eff}/Q^\star_{\rm D} = 1$,
$M_{\rm lr} = 0.5$
in Fig.~\ref{Mlr_Qeff}).

\begin{figure}
\centering
\includegraphics[width=8.5cm]{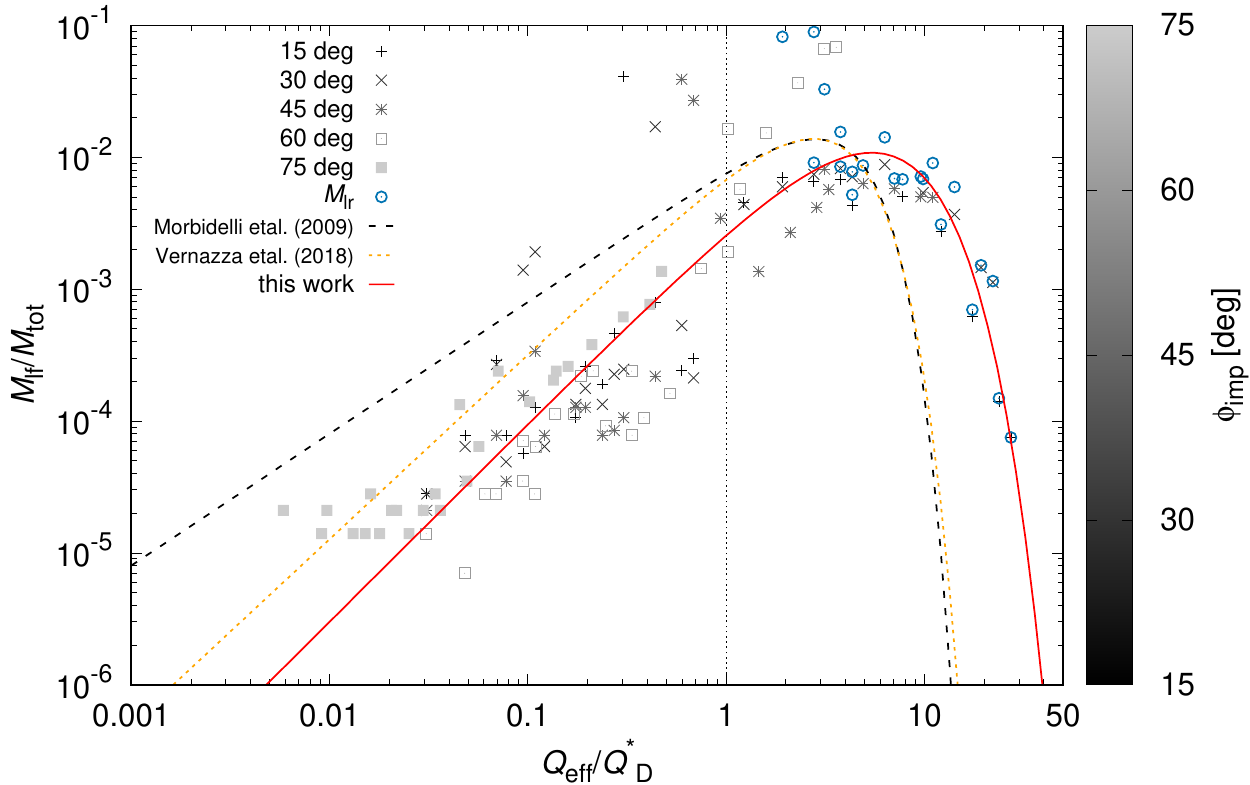}
\caption{Mass of the largest fragment~$M_{\rm lf}$ vs.\
the effective strength~$Q_{\rm eff}$.
Eq.~(\ref{Mlf}) is plotted as a solid red line.
The largest remnant mass $M_{\rm lr}$ is also plotted,
for various impact angles,
as it subsequently `disappears' for high-energy impacts.
For comparison, there are parametric relations of
\cite{Morbidelli_2009Icar..204..558M},
\cite{Vernazza_2018A&A...618A.154V}.
}
\label{Mlf_Qeff}
\end{figure}

\begin{figure}
\centering
\includegraphics[width=8.5cm]{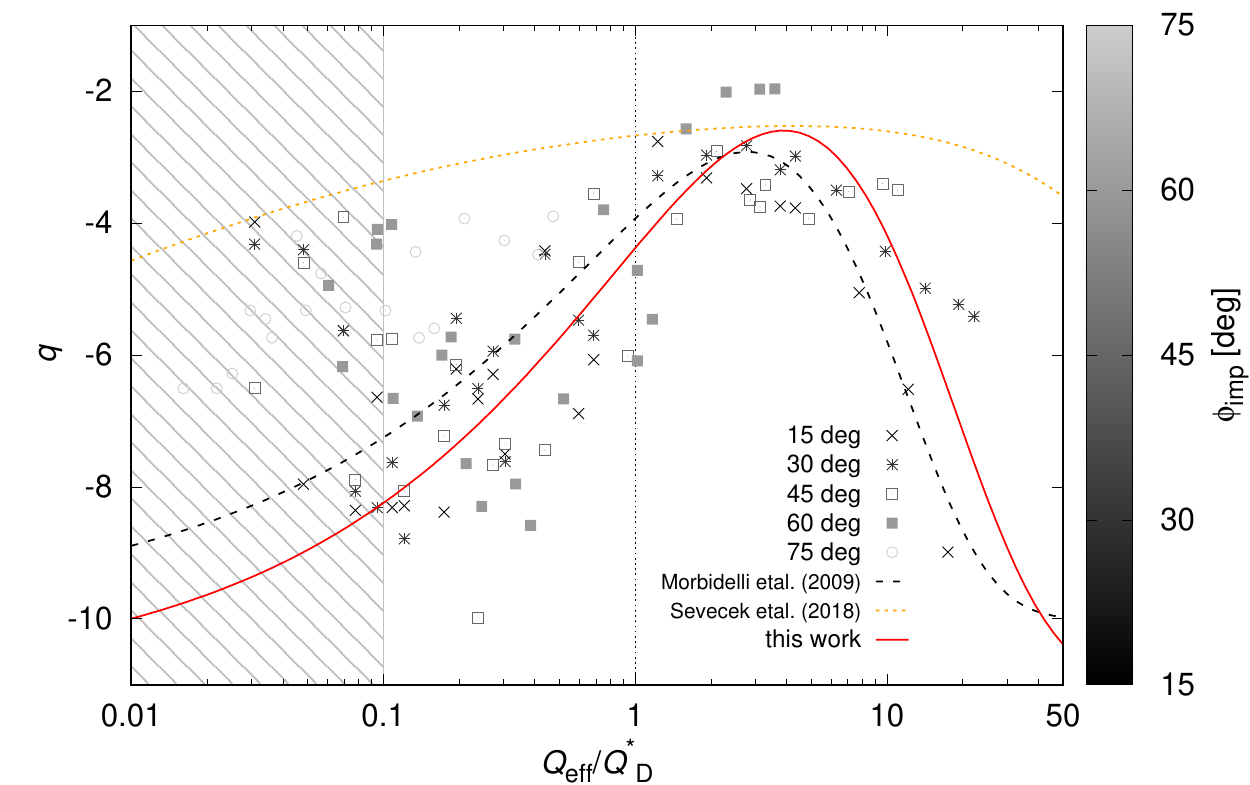}
\caption{Power-law slope of the size-frequency distribution $q$ vs.\
the effective strength~$Q_{\rm eff}$.
Eq.~(\ref{q}) is plotted as a solid red line.
There is a large scatter for low-energy impacts though
due to limited resolution (hatched area);
this part was removed from the fit.
which not taken into 
Parametric relations of
\cite{Morbidelli_2009Icar..204..558M},
\cite{Sevecek_2017Icar..296..239S}
are also shown.
}
\label{q_Qeff}
\end{figure}

Using all our SPH simulations outcomes, we derived
the following relations for the LR (Fig.~\ref{Mlr_Qeff}):
\begin{equation}
\frac{M_{\rm lr}(Q_{\rm eff})}{M_{\rm tot}} = \begin{cases}
1 - 0.2\,Q_{\rm eff}/Q^\star_{\rm D} & \hbox{for } Q_{\rm eff}/Q^\star_{\rm D} < 0.2\,, \\
0.96 - {0.46\over0.8} (Q_{\rm eff}/Q^\star_{\rm D}-0.2) & \hbox{for }   Q_{\rm eff}/Q^\star_{\rm D} < 1\,, \\
0.5 - 0.15\,(Q_{\rm eff}/Q^\star_{\rm D}-1.0) & \hbox{for } Q_{\rm eff}/Q^\star_{\rm D} \ge 1\,;\label{Mlr}\\
\end{cases}
\end{equation}
for the LF (Fig.~\ref{Mlf_Qeff}):
\begin{equation}
\frac{M_{\rm lf}(Q_{\rm eff})}{M_{\rm tot}} = 0.003\left(\frac{Q_{\rm eff}}{Q^\star_{\rm D}}\right)^{\!1.5}\!\exp\left[-\left(\frac{Q_{\rm eff}}{4.5Q^\star_{\rm D}}\right)^{\!1.2\,}\right]\,;\label{Mlf}
\end{equation}
and for the slope (Fig.~\ref{q_Qeff}):
\begin{equation}
q(Q_{\rm eff}) = -11.0 + 8.0 \left(\frac{Q_{\rm eff}}{Q^\star_{\rm D}}\right)^{\!0.45}\!\exp\left[-\left(\frac{Q_{\rm eff}}{8.0 Q^\star_{\rm D}}\right)^{\!0.8}\right]\,.\label{q}
\end{equation}
These relations are approximations of our distributions,
which are more complex, and exhibit numerous 
(possibly interesting) outliers.%
\footnote{
Outliers are expected to originate late in the reaccumuation phase,
where only a few big bodies gravitationally interact,
which is a classical N-body problem, exhibiting determininistic chaos.
For example, pairs of similarly-sized largest remnants might be formed
(see, e.g., \citealt{Vokrouhlicky_2021A&A...654A..75V}).
}
Uncertainties of the numerical coefficients typically correspond
to the last decimal place.

Alternatively,
we can fit SFDs with two slopes $q_1$, $q_2$
(similarly as in \citealt{Sevecek_2017Icar..296..239S}),
but the scatter of $q$ values is even larger
and the fit still does not fully describe the outcomes of collisions.
Moreover, for lowest-energy impacts, we do not have enough resolution
to determine the slope; this part was removed from the fit.

It may be useful to think about a different implementation
of fragmentation than using parametric relations
(as in the Boulder code; \citealt{Morbidelli_2009Icar..204..558M}),
in order to fully account for the stochasticity of collisions.
For example, using the output SFDs directly would be an option,
i.e., picking the one which is the most similar in terms of energy
and extrapolating for lowest-energy and high-energy events.
Another extrapolation would be needed within the SFDs, for $D\to 0$.

%%%%%%%%%%%%%%%%%%%%%%%%%%%%%%%%%%%%%%%%%%%%%%%%%%%%%%%%%%%%%%%%%%%

\section{Comparison with low-speed collisions}\label{sec:comparison}

We can compare Eqs.~(\ref{Mlr}) to (\ref{q}) with the original parametric relations of \cite{Morbidelli_2009Icar..204..558M}, corresponding to low-speed collisions
(see also the respective Figs.~\ref{Mlr_Qeff} to \ref{q_Qeff}).

While $M_{\rm lr}$ seems to be very similar to previous results,
thanks to the appropriate scaling by $Q_{\rm eff}$,
$M_{\rm lf}$ exhibits systematic differences.
The former is broadly given by the impact energy (cf. the scaling law),
but the latter is a result of small-scale hydrodynamics and reaccumulation.
In particular, the peak of $M_{\rm lf}$ is shifted to even higher energies.
For cratering events and sub-catastrophic disruptions,
with the ratio of impactor specific energy $Q_{\rm eff}$ to the target strength $Q^\star_{\rm D}$
smaller than $\sim 0.4$, the LF is substantially {\em smaller\/}
by almost an order of magnitude in mass (or a factor of 2 in diameter). 

This may have important consequences, because the observability of asteroid families
is practically given by the presence of a sufficiently large LF;
if it is too small, secondary collisions are unable to sustain the SFD
for a long time and the family `disappears'.
This may at least partly explain the problem with the (excessive) number of LHB families,
outlined in \cite{Broz_2013A&A...551A.117B},
but their arguments were based on catastrophic disruptions, not craterings.

In the super-catastrophic regime ($Q_{\rm eff}/Q^\star_{\rm D} \gtrsim 10$),
our simulations show the LF (again, the LR doesn't exist anymore) is substantially bigger,
which would make the observability problem outlined in Introduction even worse,
-- such families would be more observable --
but these collisions are rare,
because big projectiles are rare in the main belt.
Nevertheless, our simulations also suggest that high-speed impacts
produce actually more $D \gtrsim 10\,{\rm km}$ fragments,
which may (temporarily) enhance the collisional cascade by secondary disruptions.

%%%%%%%%%%%%%%%%%%%%%%%%%%%%%%%%%%%%%%%%%%%%%%%%%%%%%%%%%%%%%%%%%%%%%%%%

\section{Conclusions}\label{sec:conclusionsch4}

We computed a standard set of 125 simulations of high-speed
collisions between $D = 100\,{\rm km}$ asteroids
and comets. We derived parametric relations describing the dependence of
the mass of the largest remnant $M_{\rm lr}$,
the mass of the largest fragment $M_{\rm lf}$,
and the slope~$q$ of the fragment size distribution
on the effective strength $Q_{\rm eff}$.
A~comparison to low-speed collisions \citep{Durda_2007Icar..186..498D,Morbidelli_2009Icar..204..558M} shows
that the largest remnant mass~$M_{\rm lr}$ is similar as in low-speed collisions,
due to appropriate scaling with~$Q_{\rm eff}$,
but the largest fragment mass~$M_{\rm lf}$ exhibits systematic differences
--- it is typically smaller for craterings and bigger for super-catastrophic events.
This trend does not, however, explain the non-existence of old families.

Let us finally recall that relations
for macroscopic rubble-pile bodies were derived by \cite{Benavidez_2012Icar..219...57B},
for smaller $D = 10\,{\rm km}$ targets by \cite{Sevecek_2017Icar..296..239S}. 
As a future work, we plan to use all these relations in collisional models of the late heavy bombardment,
or the main belt composed of two (or more) rheologically different populations.

%%%%%%%%%%%%%%%%%%%%%%%%%%%%%%%%%%%%%%%%%%%%%%%%%%%%%%%%%%%%%%%%%%%%%%%%

\section*{Acknowledgements}\label{sec:acknowledgements}

We thank two anonymous referees for detailed reports,
which helped us to correct and improve the manuscript.
The work of JR was supported by the Grant Agency of the Charles University (grant no.\ 1109516).
The work of MB was supported by the Grant Agency of the Czech Republic (grant no.\ 21-11058S).

%%%%%%%%%%%%%%%%%%%%%%%%%%%%%%%%%%%%%%%%%%%%%%%%%%%%%%%%%%%%%%%%%%%%%%%%

\bibliographystyle{elsarticle-harv}
\bibliography{bibliography}

\begin{thebibliography}{22}
\expandafter\ifx\csname natexlab\endcsname\relax\def\natexlab#1{#1}\fi
\providecommand{\url}[1]{\texttt{#1}}
\providecommand{\href}[2]{#2}
\providecommand{\path}[1]{#1}
\providecommand{\DOIprefix}{doi:}
\providecommand{\ArXivprefix}{arXiv:}
\providecommand{\URLprefix}{URL: }
\providecommand{\Pubmedprefix}{pmid:}
\providecommand{\doi}[1]{\href{http://dx.doi.org/#1}{\path{#1}}}
\providecommand{\Pubmed}[1]{\href{pmid:#1}{\path{#1}}}
\providecommand{\bibinfo}[2]{#2}
\ifx\xfnm\relax \def\xfnm[#1]{\unskip,\space#1}\fi
%Type = Article
\bibitem[{{Benavidez} et~al.(2018){Benavidez}, {Durda}, {Enke}, {Campo
  Bagatin}, {Richardson}, {Asphaug} and
  {Bottke}}]{Benavidez_2018Icar..304..143B}
\bibinfo{author}{{Benavidez}, P.G.}, \bibinfo{author}{{Durda}, D.D.},
  \bibinfo{author}{{Enke}, B.}, \bibinfo{author}{{Campo Bagatin}, A.},
  \bibinfo{author}{{Richardson}, D.C.}, \bibinfo{author}{{Asphaug}, E.},
  \bibinfo{author}{{Bottke}, W.F.}, \bibinfo{year}{2018}.
\newblock \bibinfo{title}{{Impact simulation in the gravity regime: Exploring
  the effects of parent body size and internal structure}}.
\newblock \bibinfo{journal}{\icarus} \bibinfo{volume}{304},
  \bibinfo{pages}{143--161}.
\newblock \DOIprefix\doi{10.1016/j.icarus.2017.05.030}.
%Type = Article
\bibitem[{{Benavidez} et~al.(2012){Benavidez}, {Durda}, {Enke}, {Bottke},
  {Nesvorn{\'y}}, {Richardson}, {Asphaug} and
  {Merline}}]{Benavidez_2012Icar..219...57B}
\bibinfo{author}{{Benavidez}, P.G.}, \bibinfo{author}{{Durda}, D.D.},
  \bibinfo{author}{{Enke}, B.L.}, \bibinfo{author}{{Bottke}, W.F.},
  \bibinfo{author}{{Nesvorn{\'y}}, D.}, \bibinfo{author}{{Richardson}, D.C.},
  \bibinfo{author}{{Asphaug}, E.}, \bibinfo{author}{{Merline}, W.J.},
  \bibinfo{year}{2012}.
\newblock \bibinfo{title}{{A comparison between rubble-pile and monolithic
  targets in impact simulations: Application to asteroid satellites and family
  size distributions}}.
\newblock \bibinfo{journal}{\icarus} \bibinfo{volume}{219},
  \bibinfo{pages}{57--76}.
\newblock \DOIprefix\doi{10.1016/j.icarus.2012.01.015}.
%Type = Inproceedings
\bibitem[{{Benz}(1990)}]{Benz_1990nmns.work..269B}
\bibinfo{author}{{Benz}, W.}, \bibinfo{year}{1990}.
\newblock \bibinfo{title}{{Smooth Particle Hydrodynamics - a Review}}, in:
  \bibinfo{editor}{{Buchler}, J.R.} (Ed.), \bibinfo{booktitle}{Numerical
  Modelling of Nonlinear Stellar Pulsations Problems and Prospects}, p.
  \bibinfo{pages}{269}.
%Type = Article
\bibitem[{{Benz} and {Asphaug}(1994)}]{Benz_1994Icar..107...98B}
\bibinfo{author}{{Benz}, W.}, \bibinfo{author}{{Asphaug}, E.},
  \bibinfo{year}{1994}.
\newblock \bibinfo{title}{{Impact Simulations with Fracture. I. Method and
  Tests}}.
\newblock \bibinfo{journal}{\icarus} \bibinfo{volume}{107},
  \bibinfo{pages}{98--116}.
\newblock \DOIprefix\doi{10.1006/icar.1994.1009}.
%Type = Article
\bibitem[{{Benz} and {Asphaug}(1999)}]{Benz_1999Icar..142....5B}
\bibinfo{author}{{Benz}, W.}, \bibinfo{author}{{Asphaug}, E.},
  \bibinfo{year}{1999}.
\newblock \bibinfo{title}{{Catastrophic Disruptions Revisited}}.
\newblock \bibinfo{journal}{\icarus} \bibinfo{volume}{142},
  \bibinfo{pages}{5--20}.
\newblock \DOIprefix\doi{10.1006/icar.1999.6204},
  \href{http://arxiv.org/abs/astro-ph/9907117}{{\tt arXiv:astro-ph/9907117}}.
%Type = Article
\bibitem[{{Bro{\v{z}}} et~al.(2013){Bro{\v{z}}}, {Morbidelli}, {Bottke},
  {Rozehnal}, {Vokrouhlick{\'y}} and {Nesvorn{\'y}}}]{Broz_2013A&A...551A.117B}
\bibinfo{author}{{Bro{\v{z}}}, M.}, \bibinfo{author}{{Morbidelli}, A.},
  \bibinfo{author}{{Bottke}, W.F.}, \bibinfo{author}{{Rozehnal}, J.},
  \bibinfo{author}{{Vokrouhlick{\'y}}, D.}, \bibinfo{author}{{Nesvorn{\'y}},
  D.}, \bibinfo{year}{2013}.
\newblock \bibinfo{title}{{Constraining the cometary flux through the asteroid
  belt during the late heavy bombardment}}.
\newblock \bibinfo{journal}{\aap} \bibinfo{volume}{551}, \bibinfo{pages}{A117}.
\newblock \DOIprefix\doi{10.1051/0004-6361/201219296},
  \href{http://arxiv.org/abs/1301.6221}{{\tt arXiv:1301.6221}}.
%Type = Article
\bibitem[{{Dohnanyi}(1969)}]{Dohnanyi_1969JGR....74.2531D}
\bibinfo{author}{{Dohnanyi}, J.S.}, \bibinfo{year}{1969}.
\newblock \bibinfo{title}{{Collisional Model of Asteroids and Their Debris}}.
\newblock \bibinfo{journal}{\jgr} \bibinfo{volume}{74},
  \bibinfo{pages}{2531--2554}.
\newblock \DOIprefix\doi{10.1029/JB074i010p02531}.
%Type = Article
\bibitem[{{Durda} et~al.(2007){Durda}, {Bottke}, {Nesvorn{\'y}}, {Enke},
  {Merline}, {Asphaug} and {Richardson}}]{Durda_2007Icar..186..498D}
\bibinfo{author}{{Durda}, D.D.}, \bibinfo{author}{{Bottke}, W.F.},
  \bibinfo{author}{{Nesvorn{\'y}}, D.}, \bibinfo{author}{{Enke}, B.L.},
  \bibinfo{author}{{Merline}, W.J.}, \bibinfo{author}{{Asphaug}, E.},
  \bibinfo{author}{{Richardson}, D.C.}, \bibinfo{year}{2007}.
\newblock \bibinfo{title}{{Size-frequency distributions of fragments from SPH/
  N-body simulations of asteroid impacts: Comparison with observed asteroid
  families}}.
\newblock \bibinfo{journal}{\icarus} \bibinfo{volume}{186},
  \bibinfo{pages}{498--516}.
\newblock \DOIprefix\doi{10.1016/j.icarus.2006.09.013}.
%Type = Article
\bibitem[{{Gomes} et~al.(2005){Gomes}, {Levison}, {Tsiganis} and
  {Morbidelli}}]{Gomes_2005Natur.435..466G}
\bibinfo{author}{{Gomes}, R.}, \bibinfo{author}{{Levison}, H.F.},
  \bibinfo{author}{{Tsiganis}, K.}, \bibinfo{author}{{Morbidelli}, A.},
  \bibinfo{year}{2005}.
\newblock \bibinfo{title}{{Origin of the cataclysmic Late Heavy Bombardment
  period of the terrestrial planets}}.
\newblock \bibinfo{journal}{\nat} \bibinfo{volume}{435},
  \bibinfo{pages}{466--469}.
\newblock \DOIprefix\doi{10.1038/nature03676}.
%Type = Article
\bibitem[{{Jutzi}(2015)}]{Jutzi_2015P&SS..107....3J}
\bibinfo{author}{{Jutzi}, M.}, \bibinfo{year}{2015}.
\newblock \bibinfo{title}{{SPH calculations of asteroid disruptions: The role
  of pressure dependent failure models}}.
\newblock \bibinfo{journal}{\planss} \bibinfo{volume}{107},
  \bibinfo{pages}{3--9}.
\newblock \DOIprefix\doi{10.1016/j.pss.2014.09.012},
  \href{http://arxiv.org/abs/1502.01860}{{\tt arXiv:1502.01860}}.
%Type = Article
\bibitem[{{Monaghan}(1992)}]{Monaghan_1992ARA&A..30..543M}
\bibinfo{author}{{Monaghan}, J.J.}, \bibinfo{year}{1992}.
\newblock \bibinfo{title}{{Smoothed particle hydrodynamics.}}
\newblock \bibinfo{journal}{\araa} \bibinfo{volume}{30},
  \bibinfo{pages}{543--574}.
\newblock \DOIprefix\doi{10.1146/annurev.aa.30.090192.002551}.
%Type = Article
\bibitem[{{Morbidelli} et~al.(2009){Morbidelli}, {Bottke}, {Nesvorn{\'y}} and
  {Levison}}]{Morbidelli_2009Icar..204..558M}
\bibinfo{author}{{Morbidelli}, A.}, \bibinfo{author}{{Bottke}, W.F.},
  \bibinfo{author}{{Nesvorn{\'y}}, D.}, \bibinfo{author}{{Levison}, H.F.},
  \bibinfo{year}{2009}.
\newblock \bibinfo{title}{{Asteroids were born big}}.
\newblock \bibinfo{journal}{\icarus} \bibinfo{volume}{204},
  \bibinfo{pages}{558--573}.
\newblock \DOIprefix\doi{10.1016/j.icarus.2009.07.011},
  \href{http://arxiv.org/abs/0907.2512}{{\tt arXiv:0907.2512}}.
%Type = Article
\bibitem[{{Morbidelli} et~al.(2010){Morbidelli}, {Brasser}, {Gomes}, {Levison}
  and {Tsiganis}}]{Morbidelli_2010AJ....140.1391M}
\bibinfo{author}{{Morbidelli}, A.}, \bibinfo{author}{{Brasser}, R.},
  \bibinfo{author}{{Gomes}, R.}, \bibinfo{author}{{Levison}, H.F.},
  \bibinfo{author}{{Tsiganis}, K.}, \bibinfo{year}{2010}.
\newblock \bibinfo{title}{{Evidence from the Asteroid Belt for a Violent Past
  Evolution of Jupiter's Orbit}}.
\newblock \bibinfo{journal}{\aj} \bibinfo{volume}{140},
  \bibinfo{pages}{1391--1401}.
\newblock \DOIprefix\doi{10.1088/0004-6256/140/5/1391},
  \href{http://arxiv.org/abs/1009.1521}{{\tt arXiv:1009.1521}}.
%Type = Article
\bibitem[{{Morbidelli} et~al.(2005){Morbidelli}, {Levison}, {Tsiganis} and
  {Gomes}}]{Morbidelli_2005Natur.435..462M}
\bibinfo{author}{{Morbidelli}, A.}, \bibinfo{author}{{Levison}, H.F.},
  \bibinfo{author}{{Tsiganis}, K.}, \bibinfo{author}{{Gomes}, R.},
  \bibinfo{year}{2005}.
\newblock \bibinfo{title}{{Chaotic capture of Jupiter's Trojan asteroids in the
  early Solar System}}.
\newblock \bibinfo{journal}{\nat} \bibinfo{volume}{435},
  \bibinfo{pages}{462--465}.
\newblock \DOIprefix\doi{10.1038/nature03540}.
%Type = Article
\bibitem[{{Nesvorn{\'y}} and {Morbidelli}(2012)}]{Nesvorny_2012AJ....144..117N}
\bibinfo{author}{{Nesvorn{\'y}}, D.}, \bibinfo{author}{{Morbidelli}, A.},
  \bibinfo{year}{2012}.
\newblock \bibinfo{title}{{Statistical Study of the Early Solar System's
  Instability with Four, Five, and Six Giant Planets}}.
\newblock \bibinfo{journal}{\aj} \bibinfo{volume}{144}, \bibinfo{pages}{117}.
\newblock \DOIprefix\doi{10.1088/0004-6256/144/4/117},
  \href{http://arxiv.org/abs/1208.2957}{{\tt arXiv:1208.2957}}.
%Type = Article
\bibitem[{{Richardson} et~al.(2000){Richardson}, {Quinn}, {Stadel} and
  {Lake}}]{Richardson_2000Icar..143...45R}
\bibinfo{author}{{Richardson}, D.C.}, \bibinfo{author}{{Quinn}, T.},
  \bibinfo{author}{{Stadel}, J.}, \bibinfo{author}{{Lake}, G.},
  \bibinfo{year}{2000}.
\newblock \bibinfo{title}{{Direct Large-Scale N-Body Simulations of
  Planetesimal Dynamics}}.
\newblock \bibinfo{journal}{\icarus} \bibinfo{volume}{143},
  \bibinfo{pages}{45--59}.
\newblock \DOIprefix\doi{10.1006/icar.1999.6243}.
%Type = Article
\bibitem[{{\v{S}eve\v{c}ek} et~al.(2019){\v{S}eve\v{c}ek}, {Bro{\v{z}}} and
  {Jutzi}}]{Sevecek_2019A&A...629A.122S}
\bibinfo{author}{{\v{S}eve\v{c}ek}, P.}, \bibinfo{author}{{Bro{\v{z}}}, M.},
  \bibinfo{author}{{Jutzi}, M.}, \bibinfo{year}{2019}.
\newblock \bibinfo{title}{{Impacts into rotating targets: angular momentum
  draining and efficient formation of synthetic families}}.
\newblock \bibinfo{journal}{\aap} \bibinfo{volume}{629}, \bibinfo{pages}{A122}.
\newblock \DOIprefix\doi{10.1051/0004-6361/201935690},
  \href{http://arxiv.org/abs/1908.03248}{{\tt arXiv:1908.03248}}.
%Type = Article
\bibitem[{{\v{S}eve\v{c}ek} et~al.(2017){\v{S}eve\v{c}ek}, {Bro{\v{z}}},
  {Nesvorn{\'y}}, {Enke}, {Durda}, {Walsh} and
  {Richardson}}]{Sevecek_2017Icar..296..239S}
\bibinfo{author}{{\v{S}eve\v{c}ek}, P.}, \bibinfo{author}{{Bro{\v{z}}}, M.},
  \bibinfo{author}{{Nesvorn{\'y}}, D.}, \bibinfo{author}{{Enke}, B.},
  \bibinfo{author}{{Durda}, D.}, \bibinfo{author}{{Walsh}, K.},
  \bibinfo{author}{{Richardson}, D.C.}, \bibinfo{year}{2017}.
\newblock \bibinfo{title}{{SPH/N-Body simulations of small ($D = 10\,{\rm km}$)
  asteroidal breakups and improved parametric relations for Monte-Carlo
  collisional models}}.
\newblock \bibinfo{journal}{\icarus} \bibinfo{volume}{296},
  \bibinfo{pages}{239--256}.
\newblock \DOIprefix\doi{10.1016/j.icarus.2017.06.021},
  \href{http://arxiv.org/abs/1803.10666}{{\tt arXiv:1803.10666}}.
%Type = Misc
\bibitem[{{Tillotson}(1962)}]{Tillotson_1962geat.rept.3216T}
\bibinfo{author}{{Tillotson}, J.H.}, \bibinfo{year}{1962}.
\newblock \bibinfo{title}{{Metallic Equations of State For Hypervelocity
  Impact}}.
%Type = Article
\bibitem[{{Vernazza} et~al.(2018){Vernazza}, {Bro{\v{z}}}, {Drouard},
  {Hanu{\v{s}}}, {Viikinkoski}, {Marsset}, {Jorda}, {Fetick}, {Carry},
  {Marchis}, {Birlan}, {Fusco}, {Santana-Ros}, {Podlewska-Gaca}, {Jehin},
  {Ferrais}, {Bartczak}, {Dudzi{\'n}ski}, {Berthier}, {Castillo-Rogez},
  {Cipriani}, {Colas}, {Dumas}, {{\v{D}}urech}, {Kaasalainen}, {Kryszczynska},
  {Lamy}, {Le Coroller}, {Marciniak}, {Michalowski}, {Michel}, {Pajuelo},
  {Tanga}, {Vachier}, {Vigan}, {Warner}, {Witasse}, {Yang}, {Asphaug},
  {Richardson}, {{\v{S}}eve{\v{c}}ek}, {Gillon} and
  {Benkhaldoun}}]{Vernazza_2018A&A...618A.154V}
\bibinfo{author}{{Vernazza}, P.}, \bibinfo{author}{{Bro{\v{z}}}, M.},
  \bibinfo{author}{{Drouard}, A.}, \bibinfo{author}{{Hanu{\v{s}}}, J.},
  \bibinfo{author}{{Viikinkoski}, M.}, \bibinfo{author}{{Marsset}, M.},
  \bibinfo{author}{{Jorda}, L.}, \bibinfo{author}{{Fetick}, R.},
  \bibinfo{author}{{Carry}, B.}, \bibinfo{author}{{Marchis}, F.},
  \bibinfo{author}{{Birlan}, M.}, \bibinfo{author}{{Fusco}, T.},
  \bibinfo{author}{{Santana-Ros}, T.}, \bibinfo{author}{{Podlewska-Gaca}, E.},
  \bibinfo{author}{{Jehin}, E.}, \bibinfo{author}{{Ferrais}, M.},
  \bibinfo{author}{{Bartczak}, P.}, \bibinfo{author}{{Dudzi{\'n}ski}, G.},
  \bibinfo{author}{{Berthier}, J.}, \bibinfo{author}{{Castillo-Rogez}, J.},
  \bibinfo{author}{{Cipriani}, F.}, \bibinfo{author}{{Colas}, F.},
  \bibinfo{author}{{Dumas}, C.}, \bibinfo{author}{{{\v{D}}urech}, J.},
  \bibinfo{author}{{Kaasalainen}, M.}, \bibinfo{author}{{Kryszczynska}, A.},
  \bibinfo{author}{{Lamy}, P.}, \bibinfo{author}{{Le Coroller}, H.},
  \bibinfo{author}{{Marciniak}, A.}, \bibinfo{author}{{Michalowski}, T.},
  \bibinfo{author}{{Michel}, P.}, \bibinfo{author}{{Pajuelo}, M.},
  \bibinfo{author}{{Tanga}, P.}, \bibinfo{author}{{Vachier}, F.},
  \bibinfo{author}{{Vigan}, A.}, \bibinfo{author}{{Warner}, B.},
  \bibinfo{author}{{Witasse}, O.}, \bibinfo{author}{{Yang}, B.},
  \bibinfo{author}{{Asphaug}, E.}, \bibinfo{author}{{Richardson}, D.C.},
  \bibinfo{author}{{{\v{S}}eve{\v{c}}ek}, P.}, \bibinfo{author}{{Gillon}, M.},
  \bibinfo{author}{{Benkhaldoun}, Z.}, \bibinfo{year}{2018}.
\newblock \bibinfo{title}{{The impact crater at the origin of the Julia family
  detected with VLT/SPHERE?}}
\newblock \bibinfo{journal}{\aap} \bibinfo{volume}{618}, \bibinfo{pages}{A154}.
\newblock \DOIprefix\doi{10.1051/0004-6361/201833477}.
%Type = Article
\bibitem[{{Vokrouhlick{\'y}} et~al.(2021){Vokrouhlick{\'y}}, {Bro{\v{z}}},
  {Novakovi{\'c}} and {Nesvorn{\'y}}}]{Vokrouhlicky_2021A&A...654A..75V}
\bibinfo{author}{{Vokrouhlick{\'y}}, D.}, \bibinfo{author}{{Bro{\v{z}}}, M.},
  \bibinfo{author}{{Novakovi{\'c}}, B.}, \bibinfo{author}{{Nesvorn{\'y}}, D.},
  \bibinfo{year}{2021}.
\newblock \bibinfo{title}{{The young Hobson family: Possible binary parent body
  and low-velocity dispersal}}.
\newblock \bibinfo{journal}{\aap} \bibinfo{volume}{654}, \bibinfo{pages}{A75}.
\newblock \DOIprefix\doi{10.1051/0004-6361/202141691},
  \href{http://arxiv.org/abs/2108.05260}{{\tt arXiv:2108.05260}}.
%Type = Article
\bibitem[{{Vokrouhlick{\'y}} et~al.(2008){Vokrouhlick{\'y}}, {Nesvorn{\'y}} and
  {Levison}}]{Vokrouhlicky_2008AJ....136.1463V}
\bibinfo{author}{{Vokrouhlick{\'y}}, D.}, \bibinfo{author}{{Nesvorn{\'y}}, D.},
  \bibinfo{author}{{Levison}, H.F.}, \bibinfo{year}{2008}.
\newblock \bibinfo{title}{{Irregular Satellite Capture by Exchange Reactions}}.
\newblock \bibinfo{journal}{\aj} \bibinfo{volume}{136},
  \bibinfo{pages}{1463--1476}.
\newblock \DOIprefix\doi{10.1088/0004-6256/136/4/1463}.

\end{thebibliography}

%%%%%%%%%%%%%%%%%%%%%%%%%%%%%%%%%%%%%%%%%%%%%%%%%%%%%%%%%%%%%%%%%%%

% ALL FIGS WERE TEMPORARILY MOVED TO THIS SUBSECTION!

\onecolumn
\appendix

\section{Supplementary figures}\label{sec:supplement}

\begin{sidewaysfigure}
\centering
\includegraphics[width=24.5cm]{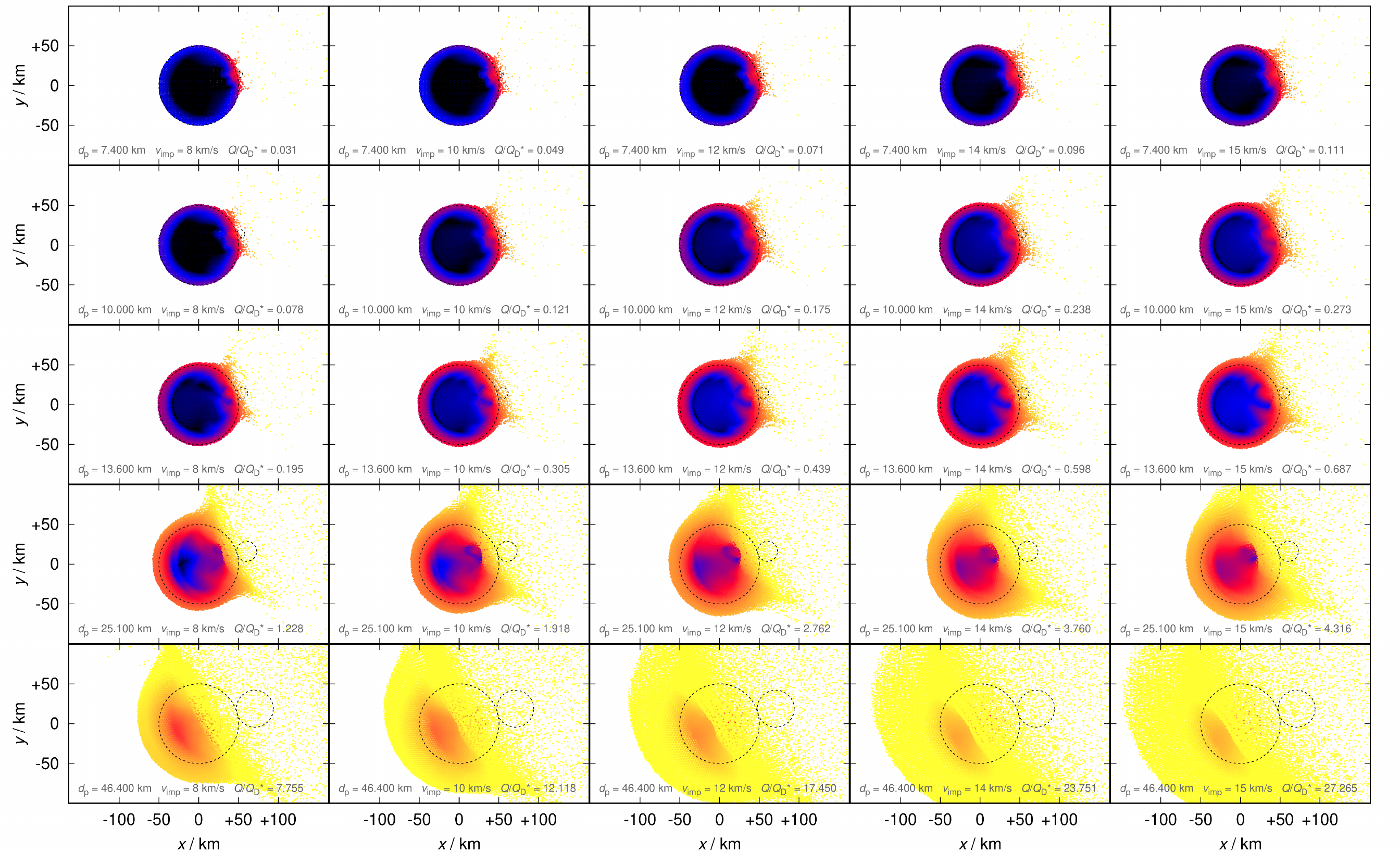}
\caption{Same as Fig.~\ref{xy_45} for the impact angle $15^\circ$.}
\label{xy_15}
\end{sidewaysfigure}

\begin{sidewaysfigure}
\centering
\includegraphics[width=24.5cm]{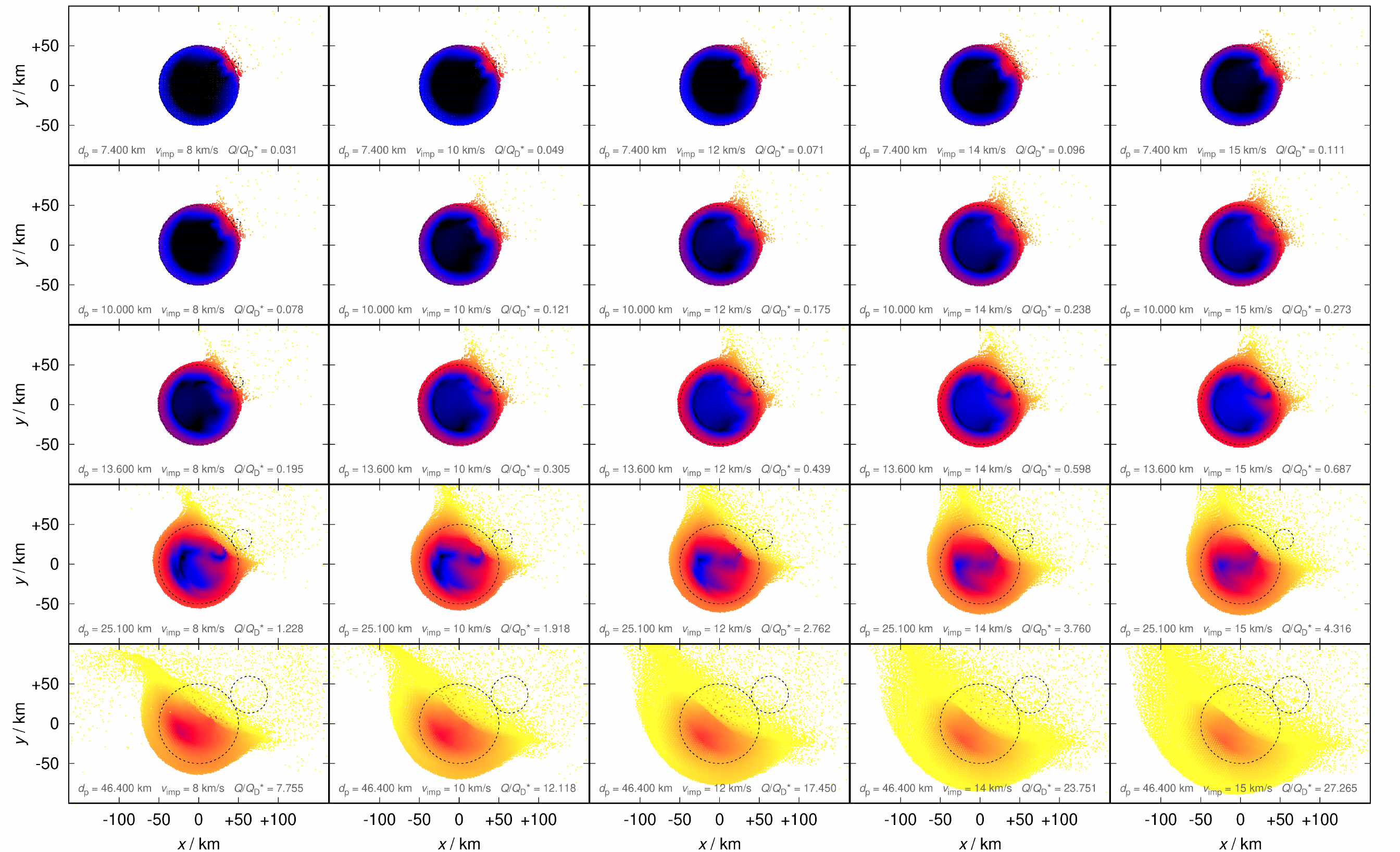}
\caption{Same as Fig.~\ref{xy_45} for the impact angle $30^\circ$.}
\label{xy_30}
\end{sidewaysfigure}

\begin{sidewaysfigure}
\centering
\includegraphics[width=24.5cm]{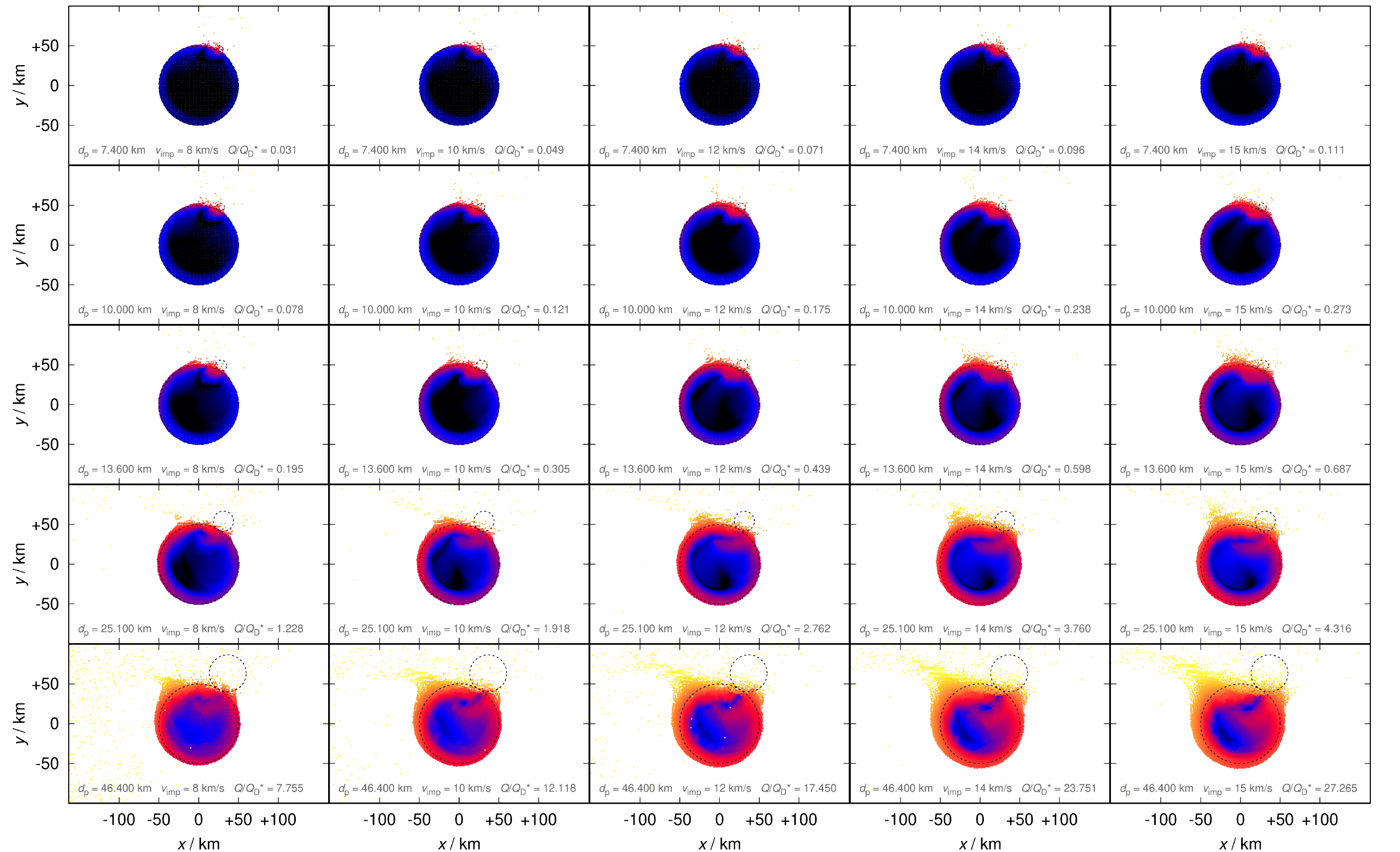}
\caption{Same as Fig.~\ref{xy_45} for the impact angle $60^\circ$.}
\label{xy_60}
\end{sidewaysfigure}

\begin{sidewaysfigure}
\centering
\includegraphics[width=24.5cm]{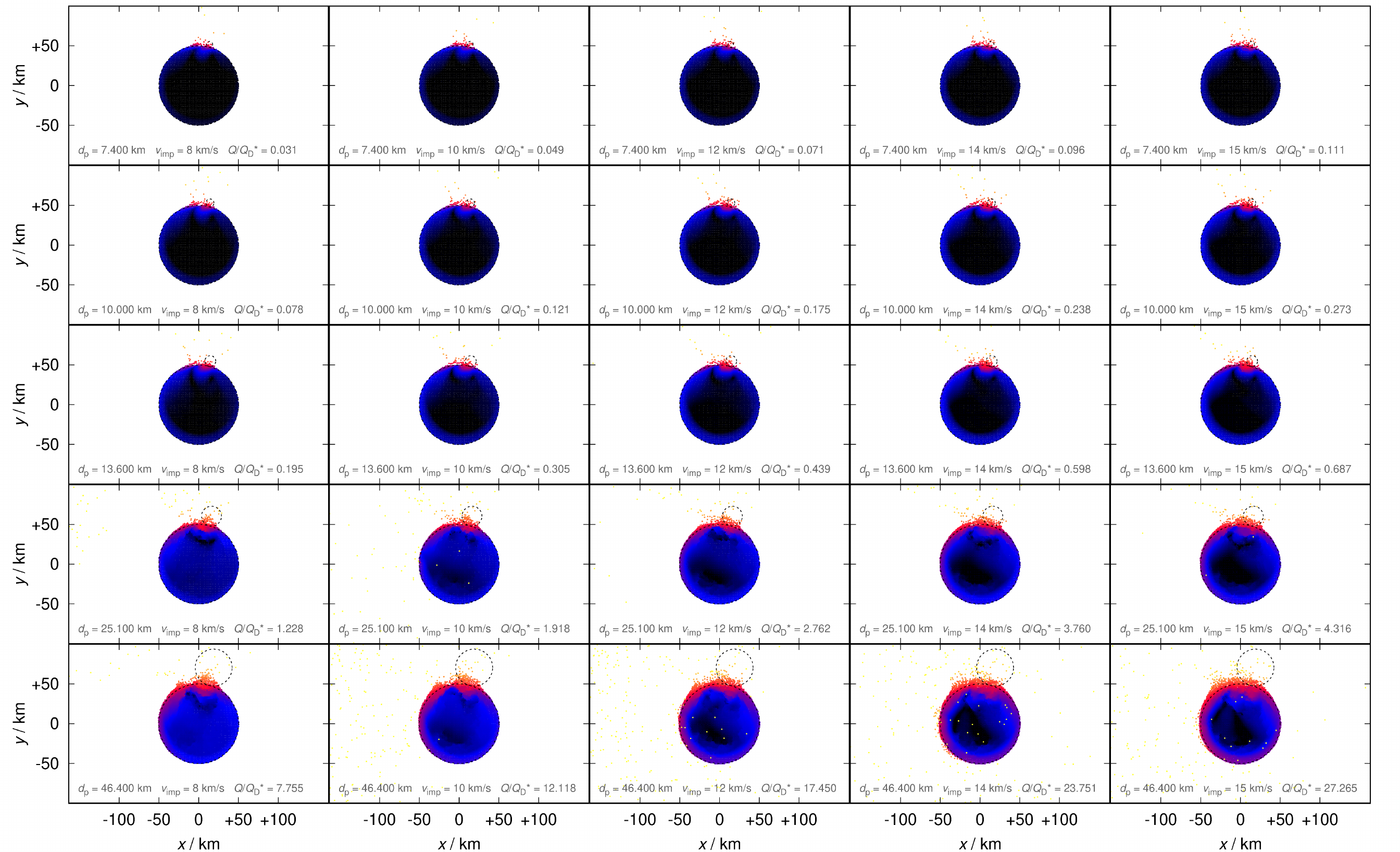}
\caption{Same as Fig.~\ref{xy_45} for the impact angle $75^\circ$.}
\label{xy_75}
\end{sidewaysfigure}

\begin{sidewaysfigure}
\centering
\includegraphics[width=24.5cm]{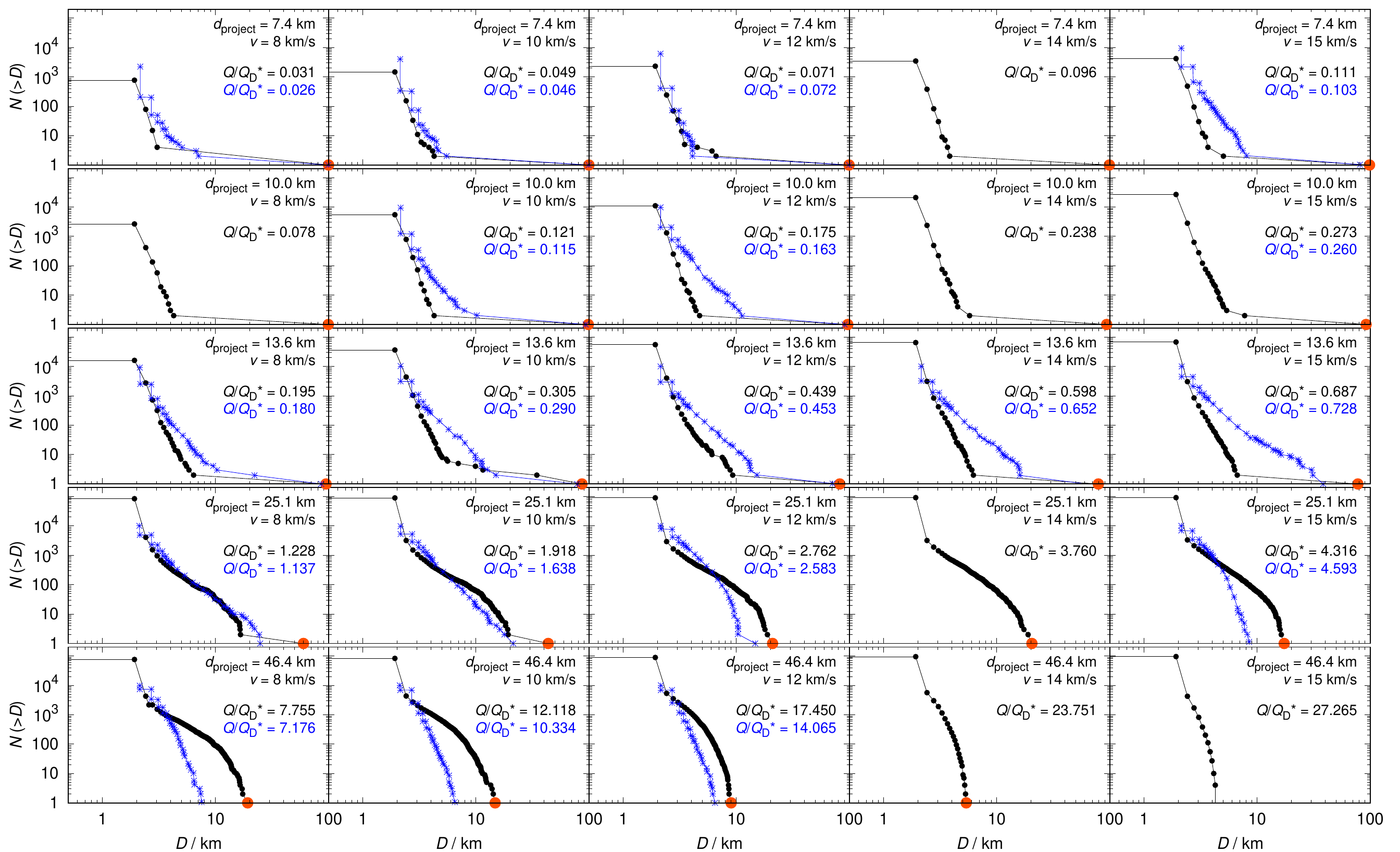}
\caption{Same as Fig.~\ref{size_distribution_45_DURDA} for the impact angle $15^\circ$.}
\label{size_distribution_15_DURDA}
\end{sidewaysfigure}

\begin{sidewaysfigure}
\centering
\includegraphics[width=24.5cm]{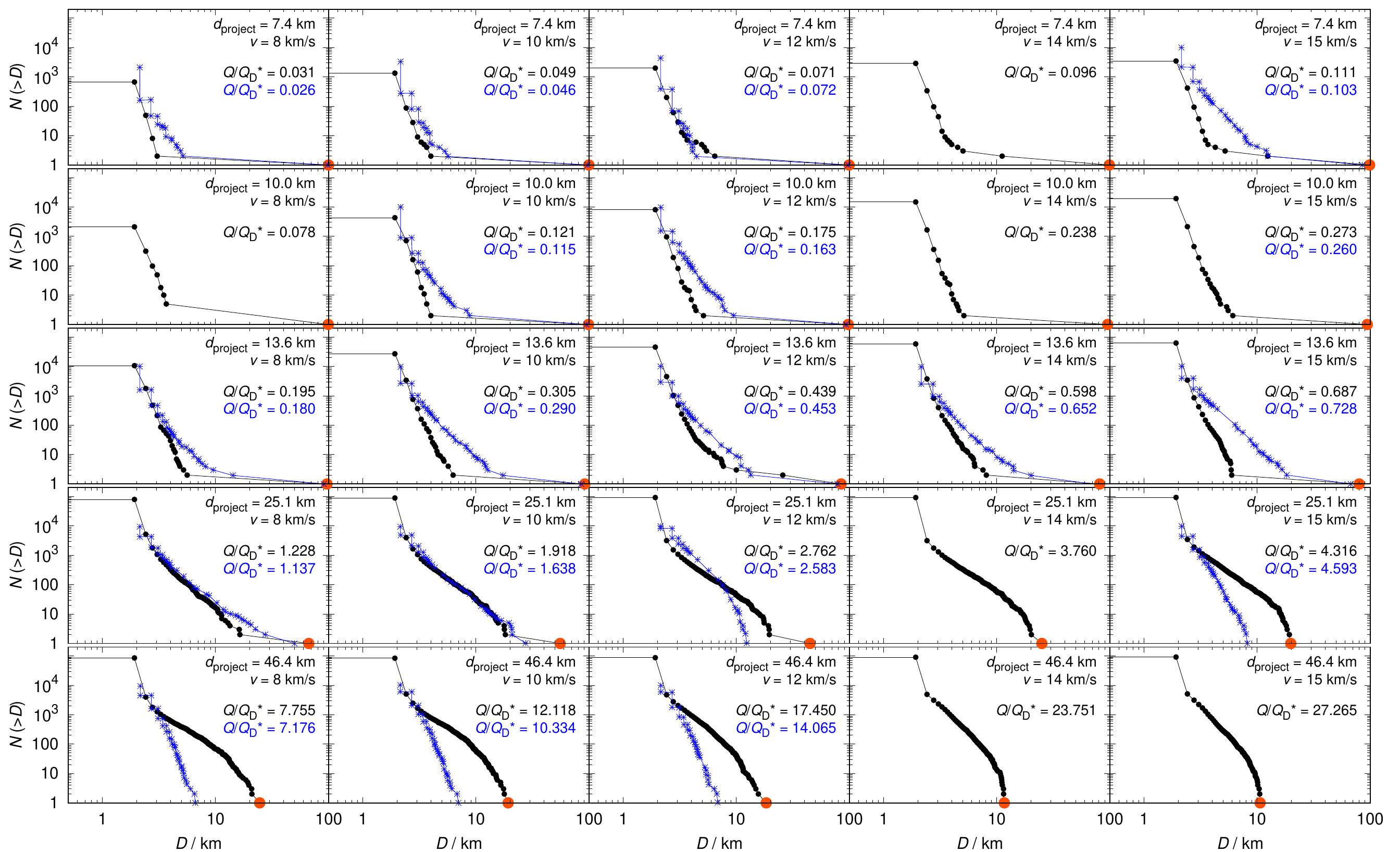}
\caption{Same as Fig.~\ref{size_distribution_45_DURDA} for the impact angle $30^\circ$.}
\label{size_distribution_30_DURDA}
\end{sidewaysfigure}

\begin{sidewaysfigure}
\centering
\includegraphics[width=24.5cm]{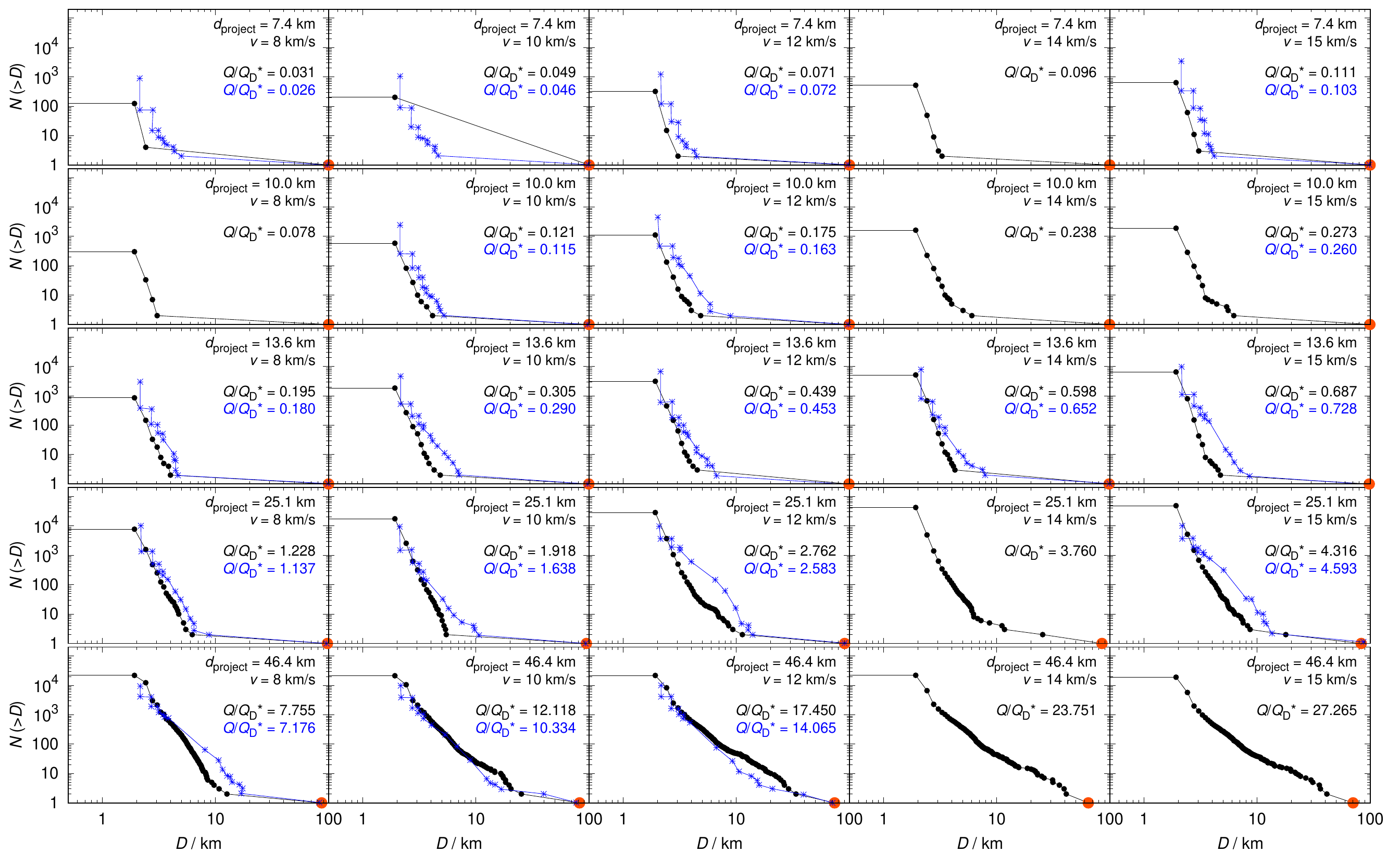}
\caption{Same as Fig.~\ref{size_distribution_45_DURDA} for the impact angle $60^\circ$.}
\label{size_distribution_60_DURDA}
\end{sidewaysfigure}

\begin{sidewaysfigure}
\centering
\includegraphics[width=24.5cm]{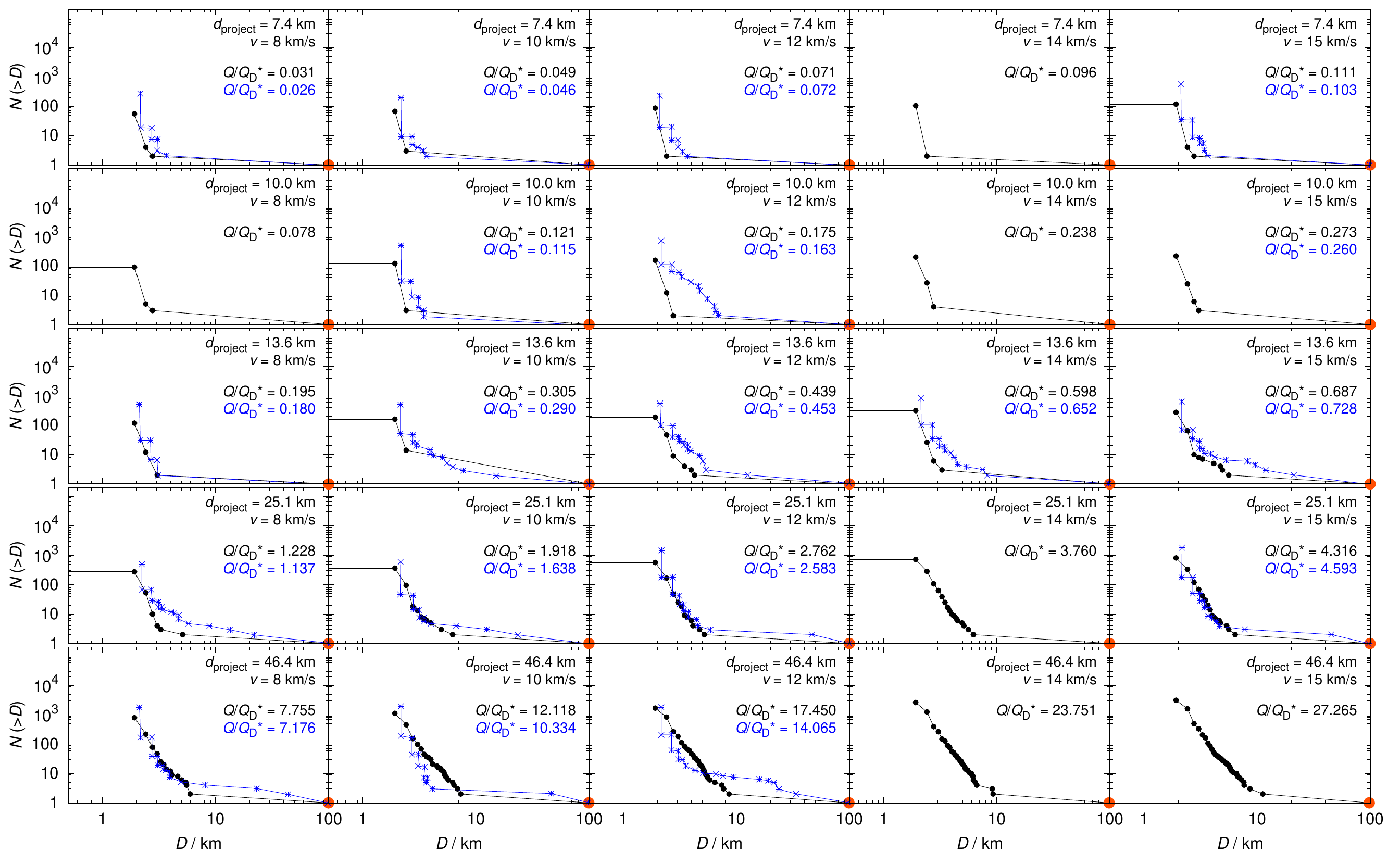}
\caption{Same as Fig.~\ref{size_distribution_45_DURDA} for the impact angle $75^\circ$.}
\label{size_distribution_75_DURDA}
\end{sidewaysfigure}

\begin{sidewaysfigure}
\centering
\includegraphics[width=24.5cm]{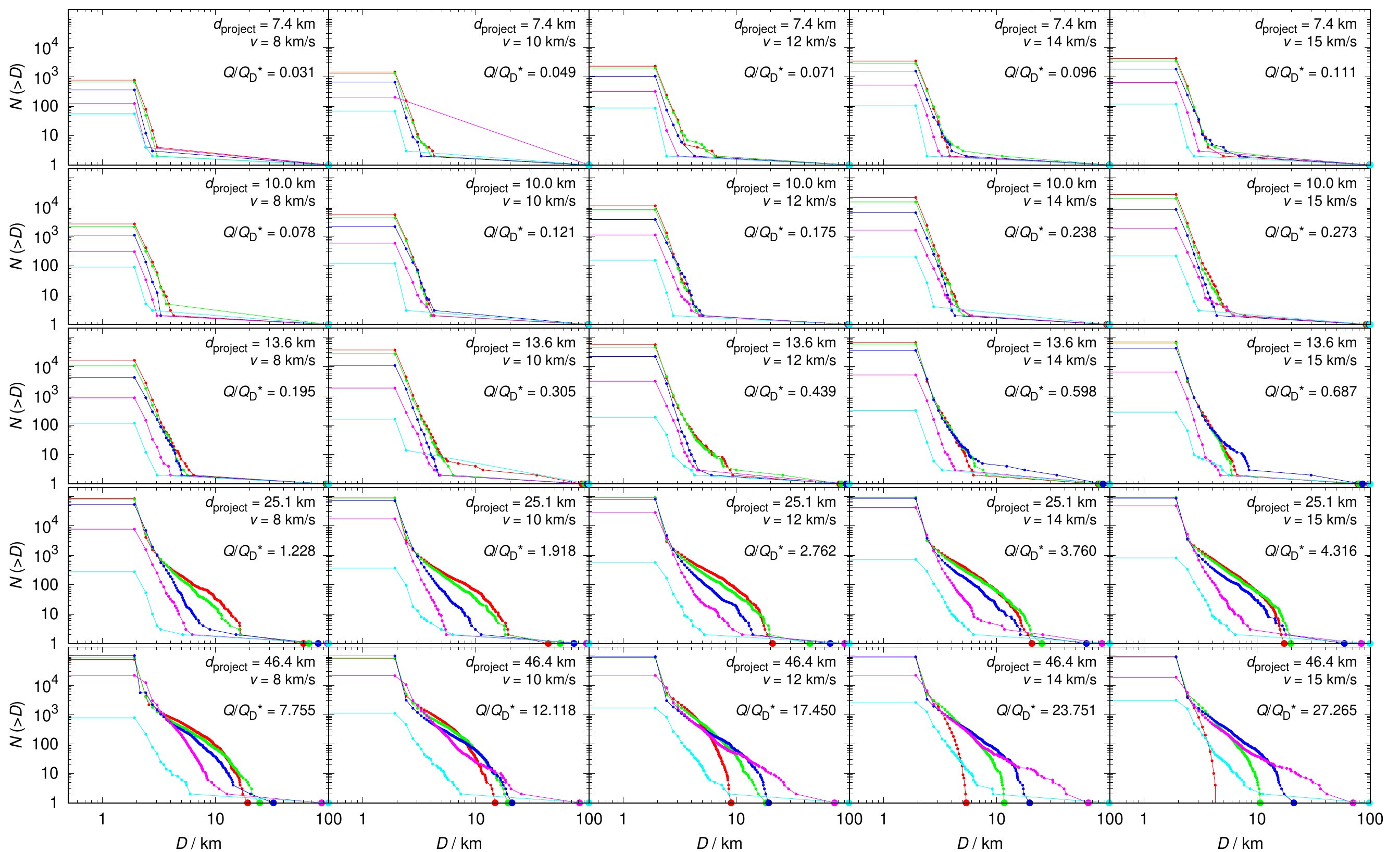}
\caption{Same as Fig.~\ref{size_distribution_45_DURDA} for all impact angles
$15^\circ$, $30^\circ$, $45^\circ$, $60^\circ$, $75^\circ$
(plotted in
\color{red}red\color{black},
\color{green}green\color{black},
\color{blue}blue\color{black},
\color{magenta}magenta\color{black},
\color{cyan}cyan\color{black}).}
\label{size_distribution_15_75}
\end{sidewaysfigure}

\begin{sidewaysfigure}
\centering
\includegraphics[width=24.5cm]{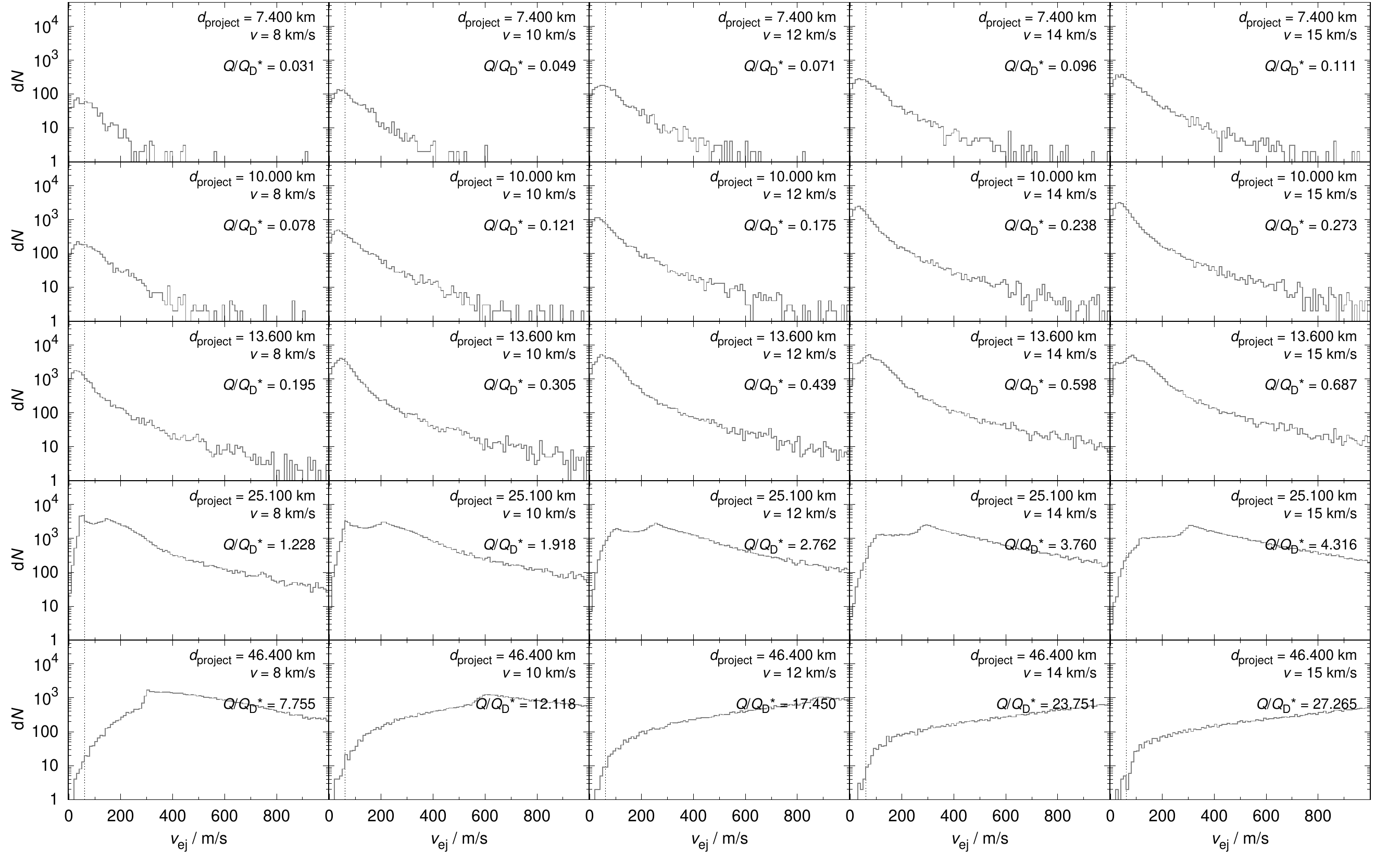}
\caption{Same as Fig.~\ref{hist_velocity_45} for the impact angle $15^\circ$.}
\label{hist_velocity_15}
\end{sidewaysfigure}

\begin{sidewaysfigure}
\centering
\includegraphics[width=24.5cm]{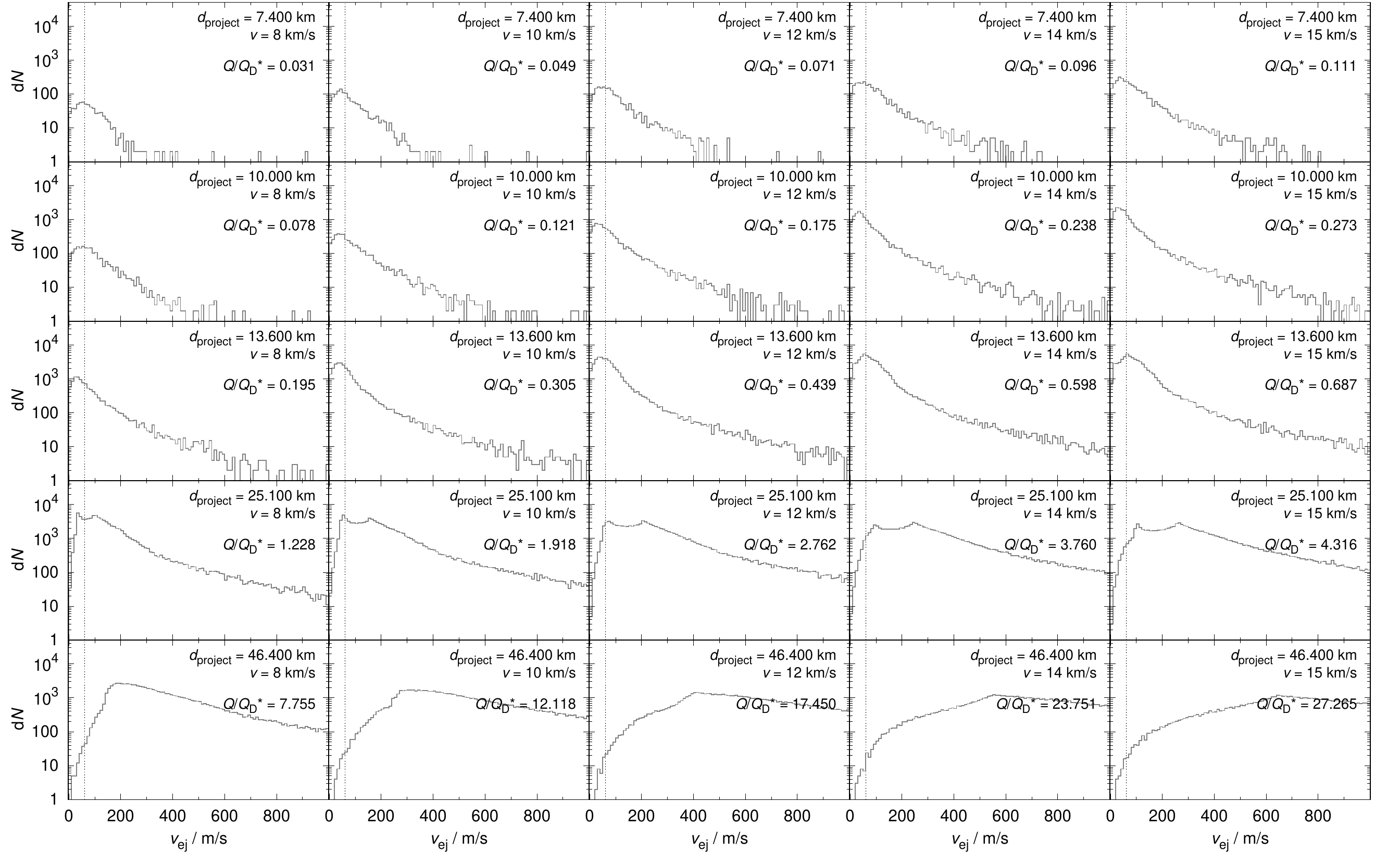}
\caption{Same as Fig.~\ref{hist_velocity_45} for the impact angle $30^\circ$.}
\label{hist_velocity_30}
\end{sidewaysfigure}

\begin{sidewaysfigure}
\centering
\includegraphics[width=24.5cm]{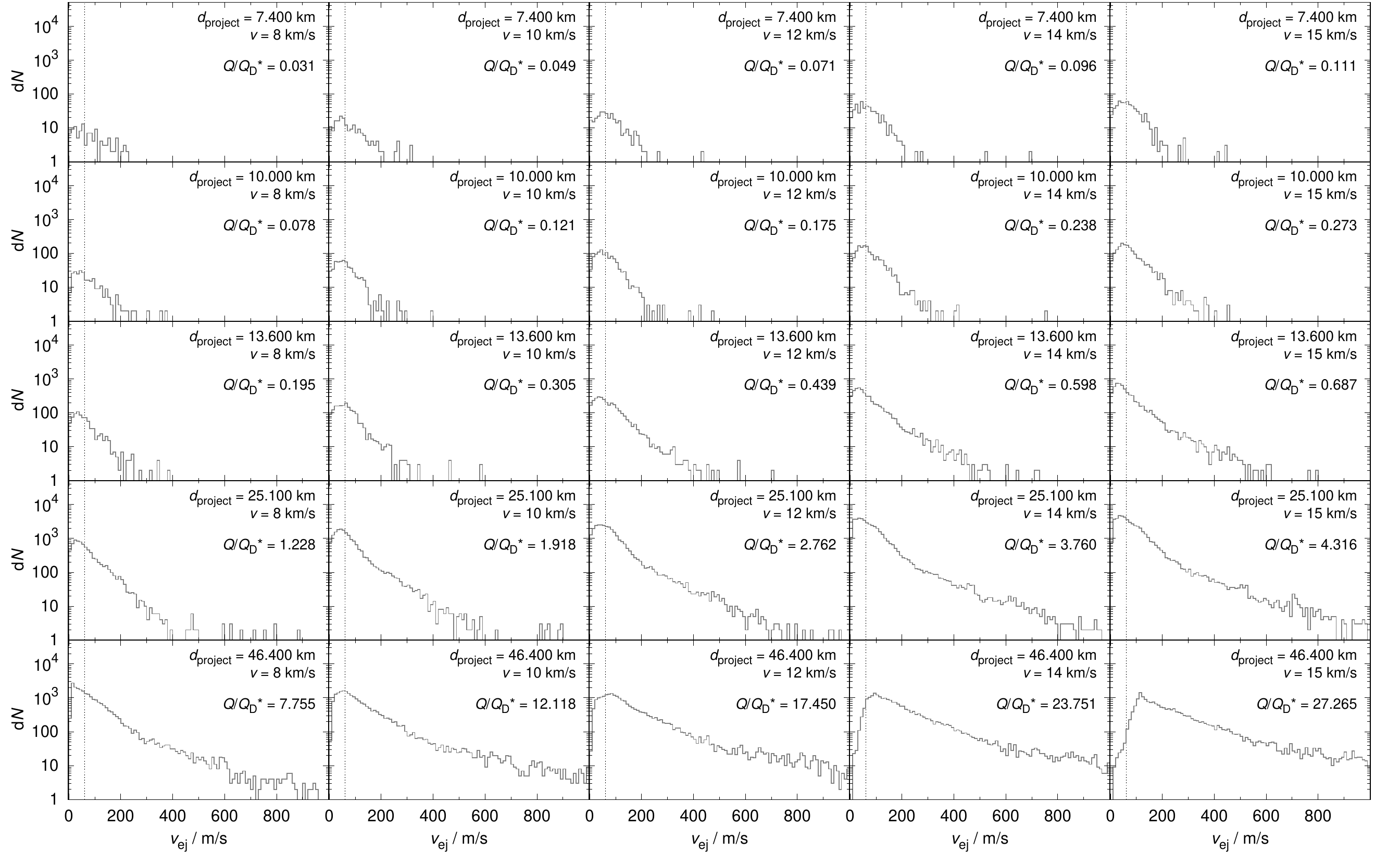}
\caption{Same as Fig.~\ref{hist_velocity_45} for the impact angle $60^\circ$.}
\label{hist_velocity_60}
\end{sidewaysfigure}

\begin{sidewaysfigure}
\centering
\includegraphics[width=24.5cm]{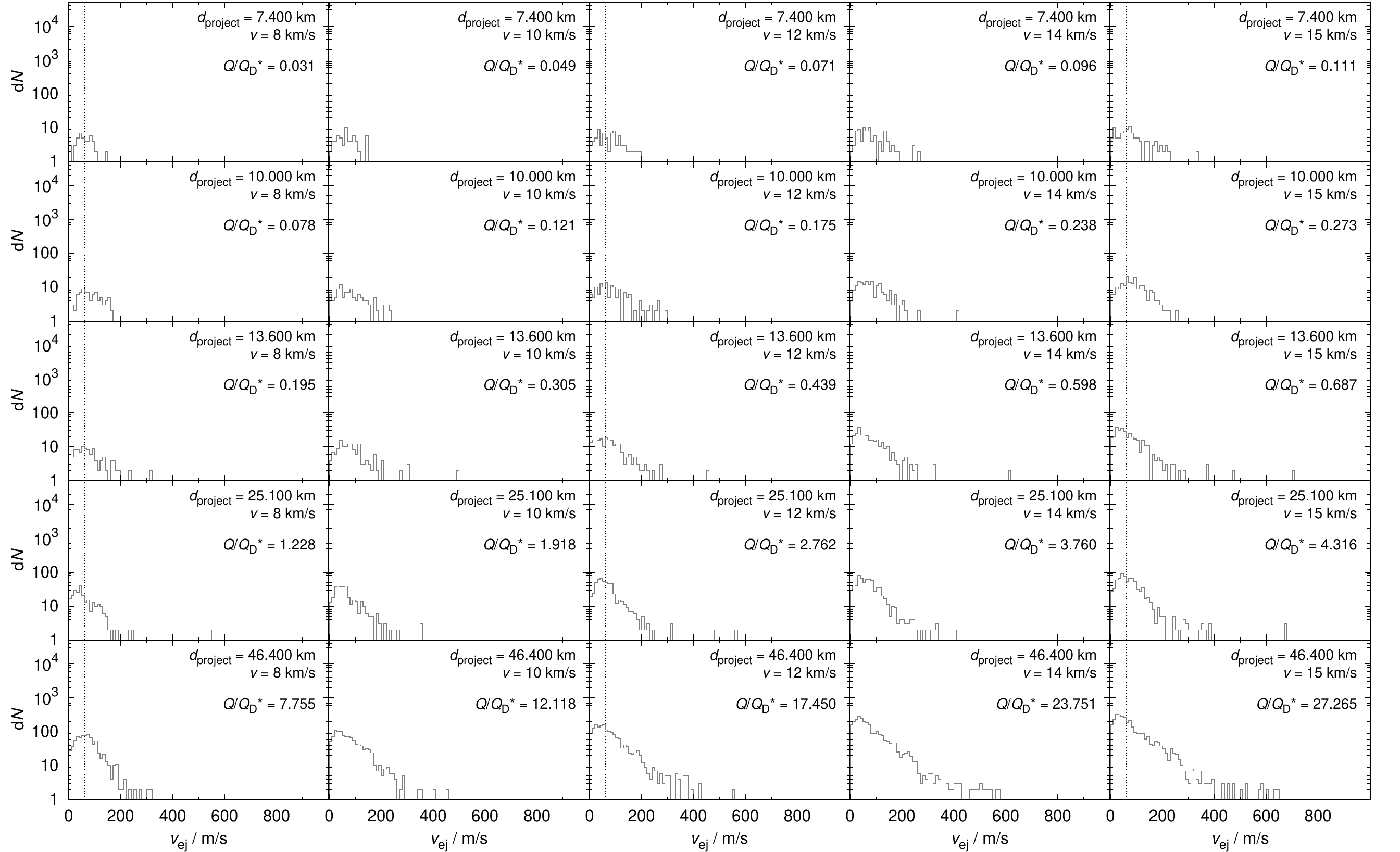}
\caption{Same as Fig.~\ref{hist_velocity_45} for the impact angle $75^\circ$.}
\label{hist_velocity_75}
\end{sidewaysfigure}

\begin{sidewaysfigure}
\centering
\includegraphics[width=24.5cm]{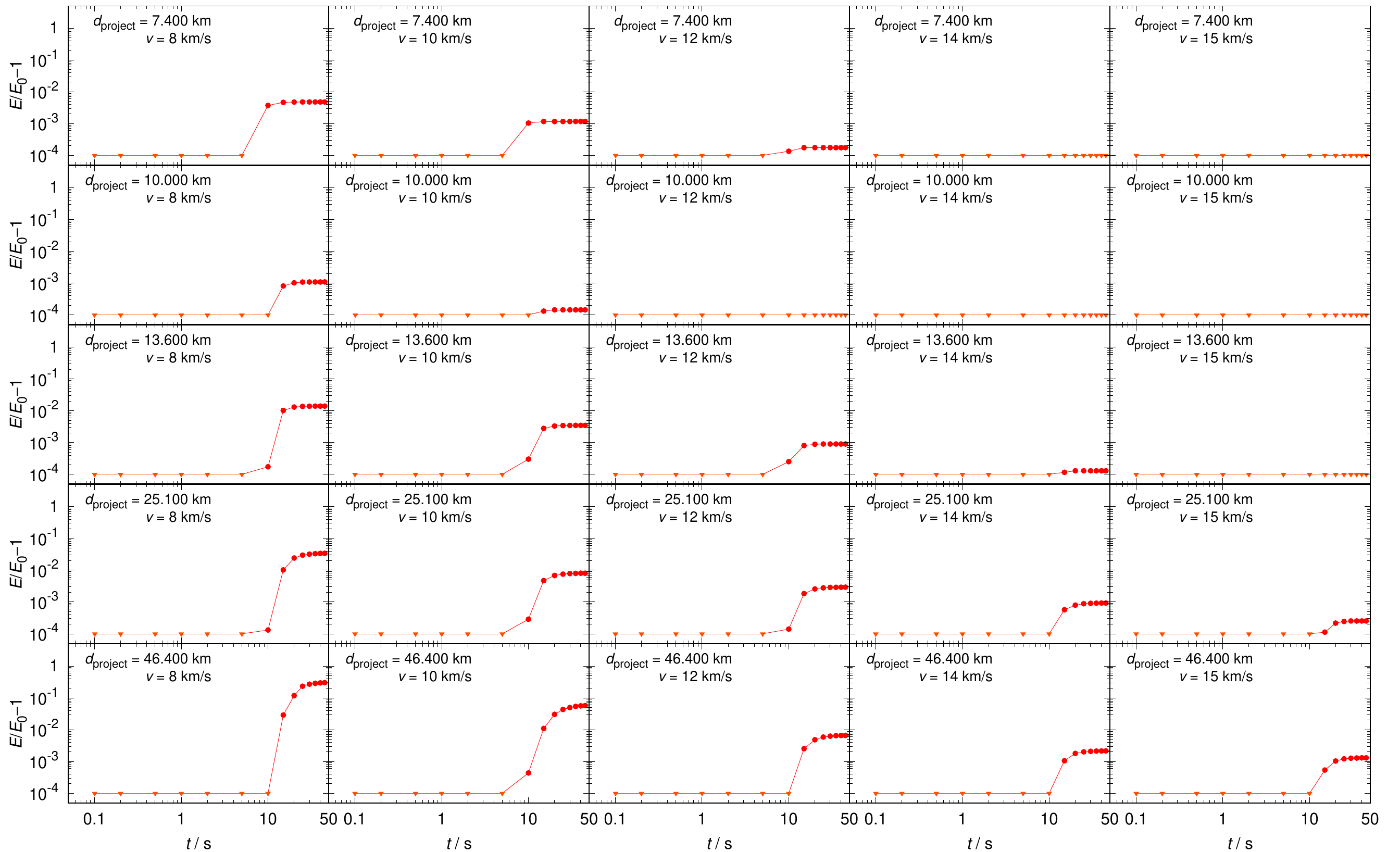}
\caption{Same as Fig.~\ref{energy_45} for the impact angle $15^\circ$.}
\label{energy_15}
\end{sidewaysfigure}

\begin{sidewaysfigure}
\centering
\includegraphics[width=24.5cm]{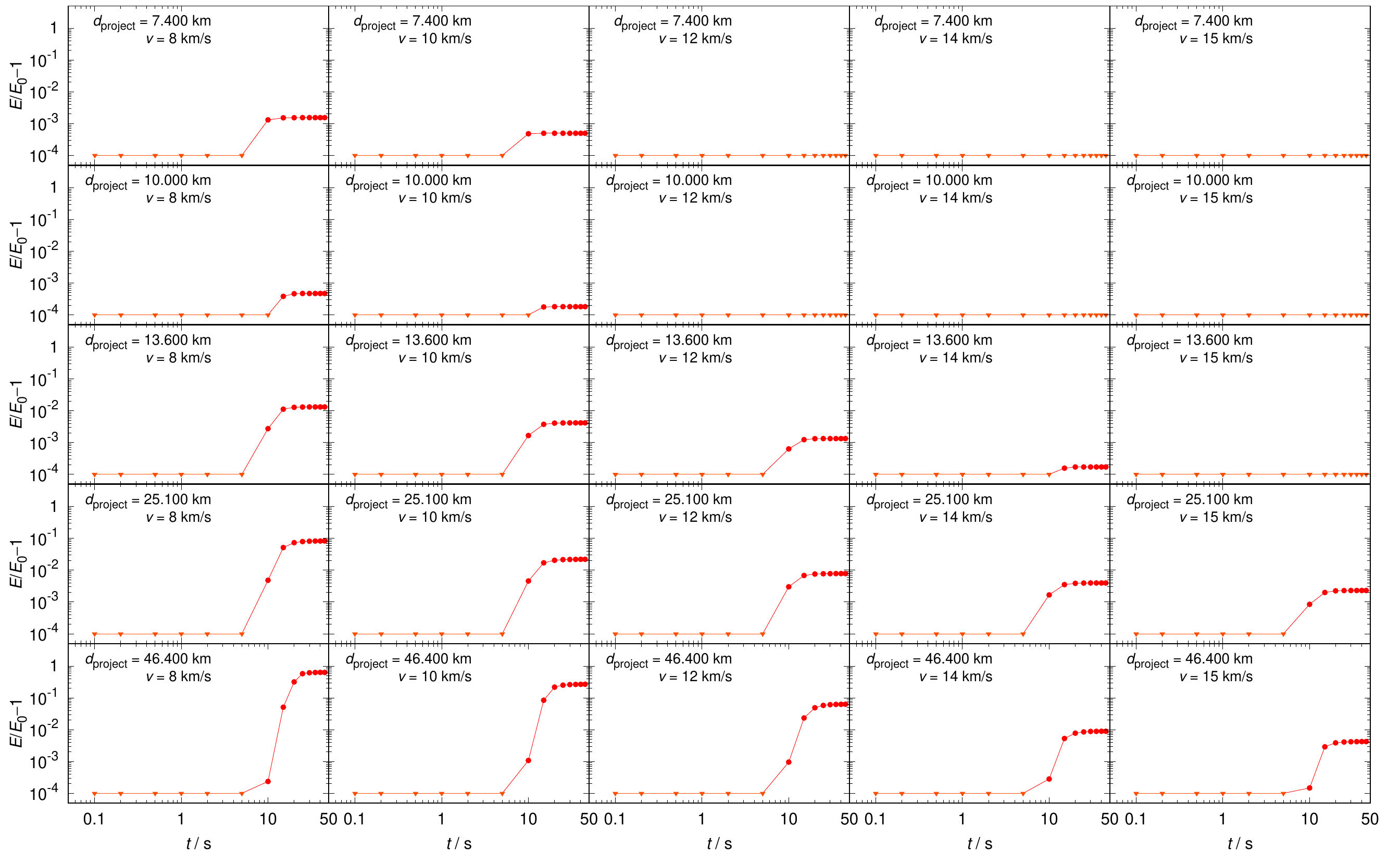}
\caption{Same as Fig.~\ref{energy_45} for the impact angle $30^\circ$.}
\label{energy_30}
\end{sidewaysfigure}

\begin{sidewaysfigure}
\centering
\includegraphics[width=24.5cm]{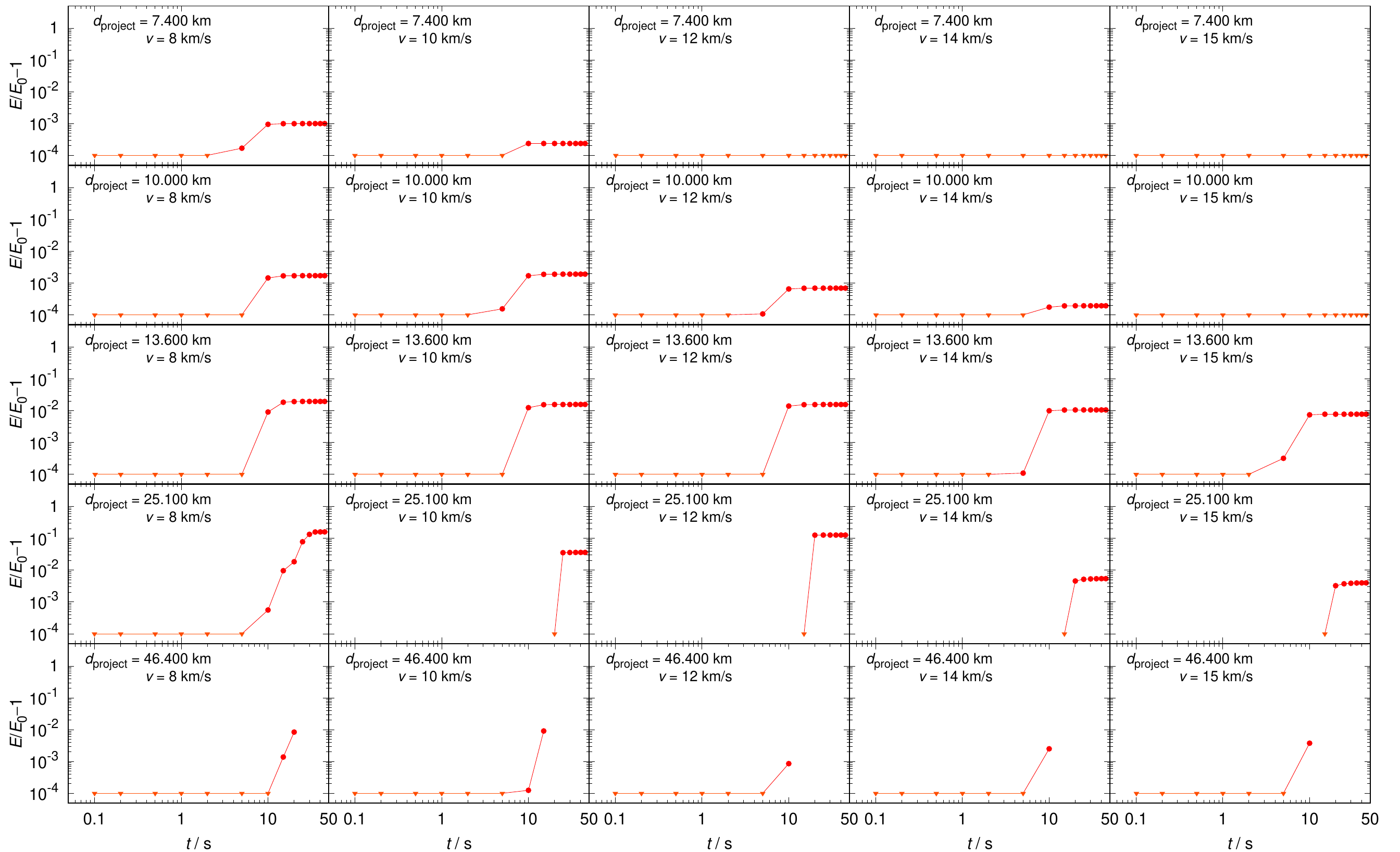}
\caption{Same as Fig.~\ref{energy_45} for the impact angle $60^\circ$.}
\label{energy_60}
\end{sidewaysfigure}

\begin{sidewaysfigure}
\centering
\includegraphics[width=24.5cm]{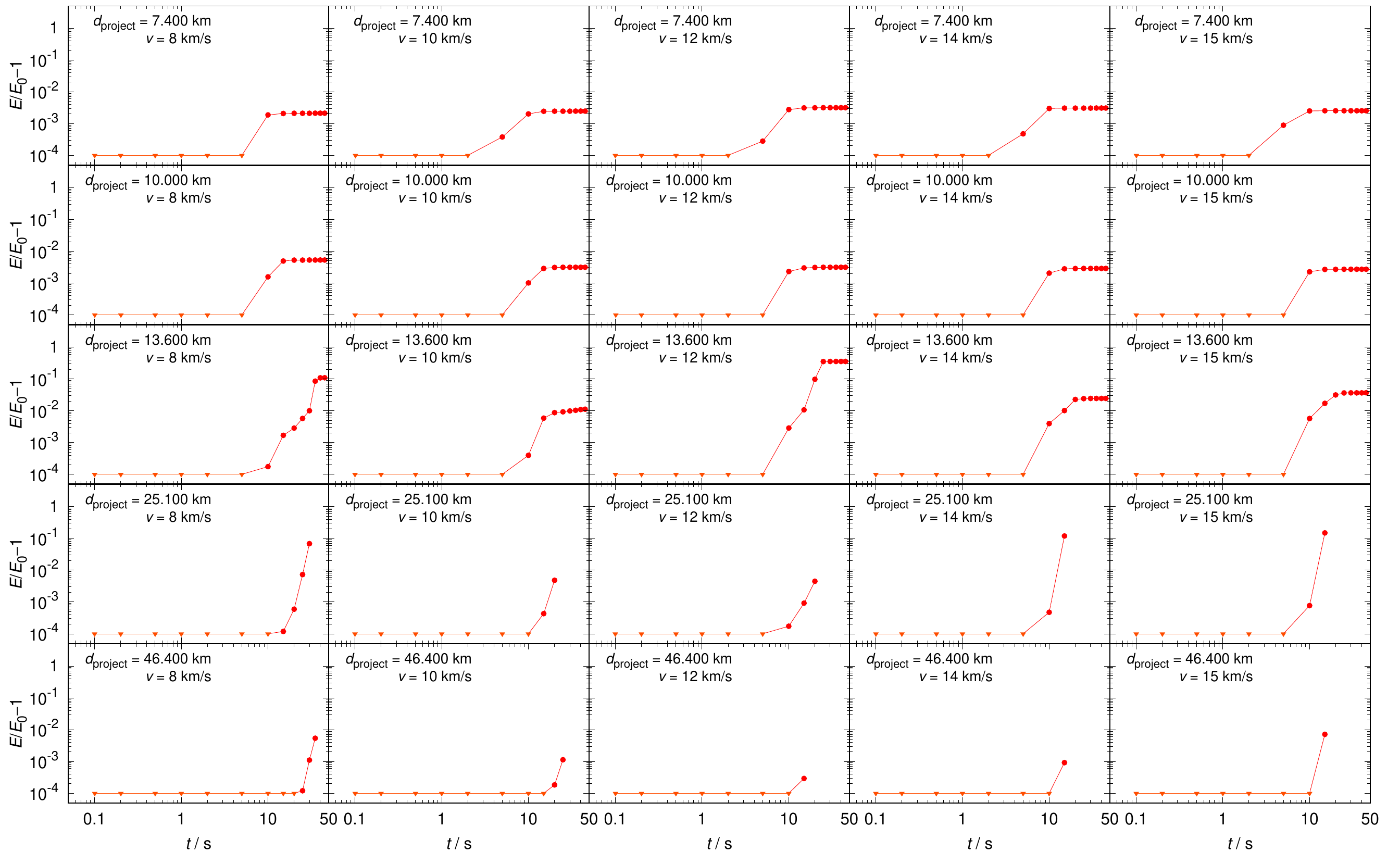}
\caption{Same as Fig.~\ref{energy_45} for the impact angle $75^\circ$.}
\label{energy_75}
\end{sidewaysfigure}

Figures for all simulations are attached:
fragmentation phase (Figs.~\ref{xy_15} to \ref{xy_75});
size-frequency distributions (Figs.~\ref{size_distribution_15_DURDA} to \ref{size_distribution_15_75});
velocity histograms (Figs.~\ref{hist_velocity_15} to \ref{hist_velocity_75});
energy vs.\ time (Figs.~\ref{energy_15} to \ref{energy_75}).

%%%%%%%%%%%%%%%%%%%%%%%%%%%%%%%%%%%%%%%%%%%%%%%%%%%%%%%%%%%%%%%%%%%

%%%%%%%%%%%%%%%%%%%%%%%%%%%%%%%%%%%%%%%%%%%%%%%%%%%%%%%%%%%%%%%%%%%

%%%%%%%%%%%%%%%%%%%%%%%%%%%%%%%%%%%%%%%%%%%%%%%%%%%%%%%%%%%%%%%%%%%

%\begin{figure}
%\centering
%\includegraphics[width=14cm]{figs4/Mlr_Q-eps-converted-to.pdf}
%\caption{Same as Fig.~\ref{Mlr_Qeff},
%but with the nominal strength~$Q$,
%instead of the effective stregth~$Q_{\rm eff}$
%(given by Eq.~(\ref{Qeff})).
%In this case, oblique impacts cannot follow the same
%dependence $M_{\rm lr}(Q)$.
%}
%\label{Mlr_Q}
%\end{figure}

%%%%%%%%%%%%%%%%%%%%%%%%%%%%%%%%%%%%%%%%%%%%%%%%%%%%%%%%%%%%%%%%%%%%%%%%

\twocolumn

\end{document}